\documentclass[preprint,showpacs,preprintnumbers,superscriptaddress,amsmath,amssymb]{revtex4}

\usepackage{graphicx}% Include figure files
\usepackage{dcolumn}% Align table columns on decimal point
\usepackage{bm,morefloats,color}% bold math

\usepackage{amssymb,bm}

\def\Eq#1{\begin{equation} #1 \end{equation}}
\def\Eqr#1{\begin{eqnarray} #1 \end{eqnarray}}
\def\Eqrsubl#1#2{\begin{subequations}\label{#1}\Eqr{#2}\end{subequations}}

\newcommand{\nn}{\nonumber}
\newcommand{\pd}{\partial}
\newcommand{\vect}[1]{\!\!\!\mbox{ \boldmath $#1$}}
\newcommand{\bea}{\begin{eqnarray}}
\newcommand{\eea}{\end{eqnarray}}

\def\Xsp{{\rm X}}

\def\Ysp{{\rm Y}}
\def\Zsp{{\rm Z}}
\def\X5sp{{\rm X}_5}
\def\Y3sp{{\rm Y}_3}
\def\Z3sp{{\rm Z}_3}

\def\Msp{{\rm M}}
\def\Nsp{{\rm N}}
\def\lap{{\triangle}}
\def\e{{\rm e}}

\begin{document}

\preprint{KU-TP 047}

\title{Cosmological intersecting brane solutions}% Force line breaks with \\

\author{Masato Minamitsuji}%
\email{masato.minamitsuji"at"kwansei.ac.jp}
\affiliation{
Graduate School of Science and Technology
Kwansei Gakuin University, Sanda 669-1337, Japan.
}%

\author{Nobuyoshi Ohta}
% \homepage{}
 \email{ohtan"at"phys.kindai.ac.jp}
%\affiliation{
%Department of Physics, Kinki University,
%Higashi-Osaka, Osaka 577-8502, Japan
%}%

\author{Kunihito Uzawa}
\affiliation{%
Department of Physics, Kinki University,
Higashi-Osaka, Osaka 577-8502, Japan
}%

\date{\today}% It is always \today, today,

\begin{abstract}
The recent discovery of an explicit dynamical description of $p$-branes
makes it possible to investigate the existence of intersection of such objects.
We generalize the solutions depending on the overall transverse space
coordinates and time to those which depend also on the relative
transverse space and satisfy new intersection rules.
We give classification of these dynamical intersecting brane solutions
involving two branes,
and discuss the application of these solutions to cosmology and show that
these give Friedmann-Lema\^itre-Robertson-Walker cosmological solutions.
Finally, we construct the brane world models, using the
(cut-)copy-paste method after compactifying the trivial spatial dimensions.
We then find that interesting brane world models
can be obtained from codimension-one branes
and several static branes with higher codimensions.
We also classify the behaviors of the brane world near
the future/past singularity.
\end{abstract}

\maketitle

%======================================%
%<<<<<<<<<<<<< SECTION 1 >>>>>>>>>>>>>>%
%======================================%

%T1>Introduction
\section{Introduction}
\label{sec:introduction}

The dynamics of the brane world model in five or six dimensions
have been much explored %\cite{Randall:1999ee}-\cite{Mukohyama:2005yw}
(see \cite{rub}-\cite{km} and references therein)
because of the possible cosmological and phenomenological interests.
Although some results have recently emerged on the applications of
the solutions in higher-dimensional supergravity
to the brane world cosmology, e.g., in 
\cite{Gibbons:2003di}-\cite{Minamitsuji:2010fp},
the construction of the 
brane world model in string theory is much less extensive.
One motivation for the present work is to improve this situation.
For this purpose, it is first necessary to construct dynamical brane
solutions depending on the time as well as space coordinates.

It has already been known~\cite{Binetruy:2007tu, Maeda:2009zi} that dynamical
brane solutions arise when the gravity is coupled not only to a 
single gauge field
but also to several combinations of scalars and forms, as generalization of
the static intersecting brane solutions in the
supergravity~\cite{Papadopoulos:1996uq}~-~\cite{Chen:2005uw}.
Similar solutions which have only time dependence have been obtained
in \cite{Ohta:2003uw} and other related solutions in
\cite{Ohta:2003rr} - \cite{Maeda:2009ds}.
Here we construct dynamical brane solutions by generalizing these
static solutions to a dynamical one.
The first class of dynamical solutions we study in this paper 
has the dependence
on the time as well as overall transverse space coordinates in the metric and
obeys the well-known intersection rules.
However, it has also been known for some time that some static intersecting brane
solutions may not follow these intersection rules 
\cite{Gauntlett:1996pb, Edelstein:1998vs, Edelstein:2004tp}.
These intersecting brane solutions are derived for the case when the branes
depend on the relative transverse directions of the intersecting branes.

%What makes possible to study these questions is that an explicit and simple
%dynamical description of branes and more general intersecting $p$-branes
%carrying brane charge has been given~\cite{Maeda:2009zi}.
%It is important to know the intersection rules for such objects,
%because we can then construct general intersecting solutions and apply these
%to cosmology and black holes.

Our goal in the present paper is to exhaust and classify 
all two-intersecting-brane
solutions which depend on the time and (relative) transverse dimensions
and to study their applications to the  cosmological evolutions and
the brane world models, in particular in the ten-dimensional string theory
and eleven-dimensional supergravity theory.
We first find cosmological solutions for possible intersections including
the above exceptional cases for two intersecting branes by extending
the similar solutions obeying the usual intersection
rules~\cite{Binetruy:2007tu, Maeda:2009zi}.
Our results on the dynamical branes are given for general cases of
arbitrary dimensions and forms,
but in their applications to cosmology and brane world models, we mainly focus on the
dynamical branes in ten- and eleven-dimensional supergravities because these are
the most important low-energy effective theories of superstrings.
We show that they exhibit physical phenomena of general interest,
including the evolution of four-dimensional universe in the
brane world cosmology and dynamics of the internal space via compactification.

The paper is organized as follows. In Sec.~\ref{sec:10D},
we show that the dynamical intersecting brane solutions of two $p$-branes exist as
an almost immediate generalization of the static $p$-brane solution.
We then go on in Sec.~\ref{sec:cosmology} to
apply these solutions to cosmology.
In Sec.~\ref{sec:app-to-bw}, we discuss construction of
brane world models from these solutions and identify physically relevant solutions
among these.
Section \ref{sec:discussions} is devoted to discussions.

%======================================%
%<<<<<<<<<<<<< SECTION 2 >>>>>>>>>>>>>>%
%======================================%

%T1>10D theory
\section{Intersecting brane solutions in $D$-dimensional theory}
\label{sec:10D}

In this section, we consider a $D$-dimensional theory
composed of the metric $g_{MN}$, dilaton $\phi$, and the antisymmetric tensor
fields of rank $(p_r+2)$ and $(p_s+2)$:
\Eqr{
S&=&\frac{1}{2\kappa^2}\int \left[R\ast{\bf 1}
 -\frac{1}{2}d\phi \wedge \ast d\phi
 -\frac{1}{2}\frac{1}{\left(p_r+2\right)!}
 \e^{\epsilon_rc_r\phi}F_{(p_r+2)}\wedge\ast F_{(p_r+2)}
 \right.\nn\\
 &&\left.
 -\frac{1}{2}\frac{1}{\left(p_s+2\right)!}
 \e^{\epsilon_sc_s\phi}F_{(p_s+2)}\wedge\ast F_{(p_s+2)}
 \right],
\label{D:action:Eq}
}
where $\kappa^2$ is the $D$-dimensional gravitational constant,
$\ast$ is the Hodge operator in the $D$-dimensional spacetime,
$F_{(n)}$ is an $n$-form field strength,
and $c_I$, $\epsilon_I~(I=r,~s)$ are constants given by
\Eqrsubl{D:parameters:Eq}{
c_I^2&=&4-\frac{2(p_I+1)(D-p_I-3)}{D-2},
   \label{D:c:Eq}\\
\epsilon_I&=&\left\{
\begin{array}{cc}
 +&~{\rm if}~~p_I-{\rm brane~is~electric}\\
 -&~~~{\rm if}~~p_I-{\rm brane~is~magnetic}
\end{array} \right.
 \label{D:epsilon:Eq}
   }
After variations with respect to the metric, the dilaton, and the forms,
we obtain the field equations:
\Eqrsubl{D:equations:Eq}{
&&R_{MN}=\frac{1}{2}\pd_M\phi \pd_N \phi
+\frac{1}{2}\frac{\e^{\epsilon_rc_r\phi}}
{\left(p_r+2\right)!}
\left[\left(p_r+2\right)
F_{MA_2\cdots A_{\left(p_r+2\right)}}
{F_N}^{A_2\cdots A_{\left(p_r+2\right)}}
-\frac{p_r+1}{D-2} g_{MN} F^2_{\left(p_r+2\right)}\right]\nn\\
&&~~~~~~+\frac{1}{2}\frac{\e^{\epsilon_sc_s\phi}}
   {\left(p_s+2\right)!}
\left[\left(p_s+2\right)
F_{MA_2\cdots A_{\left(p_s+2\right)}}
{F_N}^{A_2\cdots A_{\left(p_s+2\right)}}
-\frac{p_s+1}{D-2} g_{MN} F_{\left(p_s+2\right)}^2\right],
   \label{D:Einstein:Eq}\\
&&d\ast d\phi-\frac{1}{2}\frac{\epsilon_rc_r}
{\left(p_r+2\right)!}
\e^{\epsilon_rc_r\phi}F_{\left(p_r+2\right)}\wedge\ast F_{\left(p_r+2\right)}
-\frac{1}{2}\frac{\epsilon_sc_s}{\left(p_s+2\right)!}
\e^{\epsilon_sc_s\phi}F_{\left(p_s+2\right)}\wedge
\ast F_{\left(p_s+2\right)}=0,
   \label{D:scalar:Eq}\\
&&d\left[\e^{\epsilon_rc_r\phi}\ast F_{\left(p_r+2\right)}\right]=0,
   \label{D:gauge-r:Eq}\\
&&d\left[\e^{\epsilon_sc_s\phi}\ast F_{\left(p_s+2\right)}\right]=0.
   \label{D:gauge-s:Eq}
}

To solve these field equations, we assume that the $D$-dimensional metric
takes the form
\Eqr{
ds^2&=&h^{\alpha}_rh_s^{\beta}\left[ h_r^{-1}h_s^{-1}q_{\mu\nu}
(\Xsp)dx^{\mu}dx^{\nu}+h_s^{-1}\gamma_{ij}
(\Ysp_1)dy^idy^j\right.\nn\\
&&\left.+h_r^{-1}w_{mn}(\Ysp_2)dv^{m}dv^{n}
+u_{ab}(\Zsp)dz^adz^b\right],
 \label{D:metric:Eq}
}
where $q_{\mu\nu}$, $\gamma_{ij}$, $w_{mn}$, and $u_{ab}$ are the metrics
depending only on $x^{\mu}$, $y^i$, $v^m$, and $z^a$ coordinates of dimensions
$(p+1)$, $(p_s-p)$, $(p_r-p)$, and $(D+p-p_r-p_s-1)$, respectively.
The parameters $\alpha$ and $\beta$ in the metric
(\ref{D:metric:Eq}) are given as
\Eq{
\alpha=\frac{p_r+1}{D-2},~~~~~\beta=\frac{p_s+1}{D-2}.
 \label{D:paremeter:Eq}
}
Here we suppose that $p_s (p_r)$-brane extends along X and Y$_1$ (Y$_2$) spaces.

The $D$-dimensional metric (\ref{D:metric:Eq}) implies that
the $p$-brane solutions are characterized by a function which
depends on the coordinates transverse to the brane as well as
the worldvolume coordinate.
For the configurations of two branes, we should sort the coordinates
in three sets and the powers of harmonic functions are different for each
set of coordinates according to the intersection rules.
One set of the coordinates is the overall worldvolume coordinates, 
which are common 
to the two branes. The others are overall transverse coordinates and the
last are the relative transverse coordinates, which are transverse to only one
of the two branes.

The field equations of intersecting branes allow for the following
three kinds of possibilities on $p_r$- and $p_s$-branes
in $D$ dimensions \cite{Behrndt:1996pm, Bergshoeff:1996rn}:
\begin{description}
\item[(I)]
Both $h_r$ and $h_s$ depend on the overall transverse coordinates:
$h_r=h_r(x,z),\, h_s=h_s(x,z).$

\item[(II)]
Only $h_s$ depends on the overall transverse coordinates,
but the other $h_r$ does on the corresponding relative coordinates:
$h_r=h_r(x,y),\, h_s=h_s(x,z).$

%(And equivalently $h_r=h_r(x,z)$ and $h_s=h_s(x,v)$)

\item[(III)]
Each of $h_r$ and $h_s$ depends on the corresponding relative coordinates:
$h_r=h_r(x,y),\, h_s=h_s(x,v).$

\end{description}

In the following, we consider intersections where each participating
brane corresponds to an independent harmonic function in the solution
and derive the dynamical intersecting brane solution
in $D$ dimensions satisfying the above conditions.

%%%%%%%%%
%%%%%%%%%
%%%%%%%%%
%T2> CaseI
\subsection{Case (I)}
  \label{sub:I}

For completeness, let us first consider case (I) though
this has been already discussed in \cite{Maeda:2009zi}.
For this class, the $D$-dimensional metric~(\ref{D:metric:Eq}) becomes
\Eqr{
ds^2&=&h^{\alpha}_r(x, z)h_s^{\beta}(x, z)\left[
h_r^{-1}(x, z)h_s^{-1}(x, z)q_{\mu\nu}(\Xsp)dx^{\mu}dx^{\nu}
  +h_s^{-1}(x, z)\gamma_{ij}(\Ysp_1)dy^idy^j\right.\nn\\
&&\left. +h_r^{-1}(x, z)w_{mn}(\Ysp_2)dv^{m}dv^{n}
+u_{ab}(\Zsp)dz^adz^b\right].
 \label{1:metric:Eq}
}
We also assume that the scalar field $\phi$ and the gauge field strengths
are given as
\Eqrsubl{1:ansatz:Eq}{
\e^{\phi}&=&h_r^{\epsilon_rc_r/2}\,
h_s^{\epsilon_sc_s/2},
  \label{1:ansatz for scalar:Eq}\\
F_{\left(p_r+2\right)}&=&d\left[h^{-1}_r(x, z)\right]\wedge\Omega(\Xsp)
\wedge\Omega(\Ysp_2),
  \label{1:ansatz for gauge-r:Eq}\\
F_{\left(p_s+2\right)}&=&d\left[h^{-1}_s(x, z)\right]\wedge\Omega(\Xsp)
\wedge\Omega(\Ysp_1),
  \label{1:ansatz for gauge-s:Eq}
}
where $\Omega(\Xsp)$, $\Omega(\Ysp_1)$, and $\Omega(\Ysp_2)$
denote the volume forms of dimensions $(p+1)$, $(p_s-p)$, and $(p_r-p)$, respectively:
\Eqrsubl{1:volume:Eq}{
\Omega(\Xsp)&=&\sqrt{-q}\,dx^0\wedge dx^1\wedge \cdots \wedge
dx^p,\\
\Omega(\Ysp_1)&=&\sqrt{\gamma}\,dy^1\wedge dy^2\wedge \cdots \wedge
dy^{p_s-p},\\
\Omega(\Ysp_2)&=&\sqrt{w}\,dv^1\wedge dv^2\wedge \cdots \wedge
dv^{p_r-p}.
}
Here, $q$, $\gamma$, and $w$ are the determinant of the metric $q_{\mu\nu}$,
$\gamma_{ij}$, and $w_{mn}$, respectively.

Let us now consider gauge field equations (\ref{D:gauge-r:Eq}) and
(\ref{D:gauge-s:Eq}).
Under the assumptions (\ref{1:ansatz for gauge-r:Eq}) and
(\ref{1:ansatz for gauge-s:Eq}), we find
\Eqrsubl{1:gauge2:Eq}{
&&d\left[\e^{\epsilon_rc_r\phi}\ast F_{\left(p_r+2\right)} \right]
=-d\left[h_s^{\chi}\pd_a h_r\left(\ast_{\Zsp}dz^a\right)
\wedge\Omega(\Ysp_1)\right]=0,
  \label{1:gauge2-r:Eq}\\
&&d\left[\e^{\epsilon_sc_s\phi}\ast F_{\left(p_s+2\right)} \right]
=-d\left[h_r^{\chi}\pd_a h_s\left(\ast_{\Zsp}dz^a\right)
\wedge\Omega(\Ysp_2)\right]=0,
  \label{1:gauge2-s:Eq}
 }
where $\ast_{\Zsp}$ denotes the Hodge operator on
$\Zsp$, and $\chi$ is defined by
\Eq{
\chi=p+1-\frac{\left(p_r+1\right)\left(p_s+1\right)}{D-2}
+\frac{1}{2}\epsilon_r\epsilon_sc_rc_s.
   \label{1:chi:Eq}
}
The vanishing condition of $\chi$ is the intersection
rule~\cite{Argurio:1997gt,Ohta:1997gw,Maeda:2009zi}.
Then, Eq.~(\ref{1:gauge2-r:Eq}) leads to
\Eq{
h_s^{\chi}\lap_{\Zsp}h_r=0,~~~~\pd_{\mu}h_s^{\chi}\pd_a h_r
+h_s^{\chi}\pd_{\mu}\pd_a h_r=0,
  \label{1:gauge-r2:Eq}
}
where $\triangle_{\Zsp}$ is the Laplace operators on the space of
$\Zsp$.
On the other hand, it follows from (\ref{1:gauge2-s:Eq}) that
\Eq{
h_r^{\chi}\lap_{\Zsp} h_s=0,~~~~\pd_{\mu}h_r^{\chi}\pd_a h_s
+h_r^{\chi}\pd_{\mu}\pd_a h_s=0.
  \label{1:gauge-s2:Eq}
}
When the intersection rule $\chi=0$ is obeyed, Eq.~(\ref{1:gauge-r2:Eq}) gives
\Eq{
\lap_{\Zsp}h_r=0,~~~~\pd_{\mu}\pd_a h_r=0,
    \label{1:gauge3:Eq}
}
and Eq.~(\ref{1:gauge-s2:Eq}) reduces to
\Eq{
\lap_{\Zsp}h_s=0,~~~~\pd_{\mu}\pd_a h_s=0.
   \label{1:gauge4:Eq}
}
%{\bf
We note that, in this case, the functions $h_r$ and $h_s$ can
be written by linear combinations of terms depending on both $x^{\mu}$
and $z^a$. We are now going to see that the Einstein equations also hold
if we use this result and the intersection rule $\chi=0$.
%}

Next we consider the Einstein equations (\ref{D:Einstein:Eq}).
With the assumptions (\ref{1:metric:Eq}) and (\ref{1:ansatz:Eq}),
they reduce to
\Eqrsubl{1:cEinstein:Eq}{
&&R_{\mu\nu}(\Xsp)
-h_r^{-1}D_{\mu}D_{\nu}h_r-h_s^{-1}D_{\mu}D_{\nu}h_s
-\frac{1}{2}\left(h_rh_s\right)^{-1}\left(\pd_{\mu}h_r\pd_{\nu}h_s
+\pd_{\mu}h_s\pd_{\nu}h_r\right)\nn\\
&&~~~~-\frac{1}{2}(\alpha+\beta-2)q_{\mu\nu}q^{\rho\sigma}
\pd_{\sigma}\ln h_r\pd_{\sigma}\ln h_s
-\frac{1}{2}q_{\mu\nu}h_r^{-1}(\alpha-1)\left[\lap_{\Xsp}h_r
+\left(h_rh_s\right)^{-1}\lap_{\Zsp}h_r\right]\nn\\
&&~~~~-\frac{1}{2}q_{\mu\nu}h_s^{-1}(\beta-1)\left[\lap_{\Xsp}h_s
+\left(h_rh_s\right)^{-1}\lap_{\Zsp}h_s\right]=0,
 \label{1:cEinstein-mu:Eq}\\
&&h_r^{-1}\pd_{\mu}\pd_a h_r=0,
 \label{1:cEinstein-mi:Eq}\\
&&h_s^{-1}\pd_{\mu}\pd_a h_s=0,
 \label{1:cEinstein-mi2:Eq}\\
&&R_{ij}(\Ysp_1)-\frac{1}{2}\left(\alpha+\beta-1\right)
\gamma_{ij}h_rq^{\rho\sigma}\pd_{\rho}\ln h_r\pd_{\sigma}\ln h_s
-\frac{1}{2}\gamma_{ij}\alpha\left[\lap_{\Xsp}h_r
+\left(h_rh_s\right)^{-1}\lap_{\Zsp}h_r\right]\nn\\
&&~~~~-\frac{1}{2}\gamma_{ij}(\beta-1)h_rh_s^{-1}
\left[\lap_{\Xsp}h_s+\left(h_rh_s\right)^{-1}\lap_{\Zsp}h_s\right]=0,
 \label{1:cEinstein-ij:Eq}\\
&&R_{mn}(\Ysp_2)-\frac{1}{2}\left(\alpha+\beta-1\right)
w_{mn}h_sq^{\rho\sigma}\pd_{\rho}\ln h_r\pd_{\sigma}\ln h_s
-\frac{1}{2}\beta w_{mn}
\left[\lap_{\Xsp}h_s+\left(h_rh_s\right)^{-1}\lap_{\Zsp}h_s\right]\nn\\
&&~~~~-\frac{1}{2}(\alpha-1)w_{mn}h_r^{-1}h_s
\left[\lap_{\Xsp}h_r+\left(h_rh_s\right)^{-1}\lap_{\Zsp}h_r\right]=0,
 \label{1:cEinstein-mn:Eq}\\
&&R_{ab}(\Zsp)-\frac{1}{2}\left(\alpha+\beta\right)
u_{ab}h_rh_sq^{\rho\sigma}\pd_{\rho}\ln h_r\pd_{\sigma}\ln h_s
-\frac{1}{2}\alpha u_{ab}h_s
\left[\lap_{\Xsp}h_r+\left(h_rh_s\right)^{-1}\lap_{\Zsp}h_r\right]\nn\\
&&~~~~-\frac{1}{2}\beta u_{ab}h_r\left[\lap_{\Xsp}h_s
+\left(h_rh_s\right)^{-1}\lap_{\Zsp}h_s\right]=0,
  \label{1:cEinstein-ab:Eq}
}
where we have used the intersection rule $\chi=0$, and
$D_{\mu}$ is the covariant derivative with respect to
the metric $q_{\mu\nu}$, $\triangle_{\Xsp}$ is
the Laplace operators on the space of
$\Xsp$, and
$R_{\mu\nu}(\Xsp)$, $R_{ij}(\Ysp_1)$, $R_{mn}(\Ysp_2)$,
and $R_{ab}(\Zsp)$ are the Ricci tensors
of the metrics $q_{\mu\nu}(\Xsp)$, $\gamma_{ij}(\Ysp_1)$,
$w_{mn}(\Ysp_2)$, and $u_{ab}(\Zsp)$, respectively.

We see from Eqs.~(\ref{1:cEinstein-mi:Eq}) and (\ref{1:cEinstein-mi2:Eq})
that the warp factors $h_r$ and $h_s$ must be of the form
\Eq{
h_r(x, z)= h_0(x)+h_1(z),~~~~h_s(x, z)= k_0(x)+k_1(z).
  \label{1:warp:Eq}
}
With this form of $h_r$ and $h_s$, the other components of
the Einstein equations (\ref{1:cEinstein:Eq}) are rewritten as
\Eqrsubl{1:c2Einstein:Eq}{
&&R_{\mu\nu}(\Xsp)
-h_r^{-1}D_{\mu}D_{\nu}h_0-h_s^{-1}D_{\mu}D_{\nu}k_0
-\frac{1}{2}\left(h_rh_s\right)^{-1}\left(\pd_{\mu}h_0\pd_{\nu}k_0
+\pd_{\mu}k_0\pd_{\nu}h_0\right)\nn\\
&&~~~~-\frac{1}{2}(\alpha+\beta-2)\left(h_rh_s\right)^{-1}
q_{\mu\nu}q^{\rho\sigma}\pd_{\sigma}h_0\pd_{\sigma}k_0
-\frac{1}{2}q_{\mu\nu}h_r^{-1}(\alpha-1)\left[\lap_{\Xsp}h_0
+\left(h_rh_s\right)^{-1}\lap_{\Zsp}h_1\right]\nn\\
&&~~~~-\frac{1}{2}q_{\mu\nu}h_s^{-1}(\beta-1)
\left[\lap_{\Xsp}k_0+\left(h_rh_s\right)^{-1}\lap_{\Zsp}k_1\right]=0,
 \label{1:c2Einstein-mu:Eq}\\
&&R_{ij}(\Ysp_1)-\frac{1}{2}\left(\alpha+\beta-1\right)
\gamma_{ij}h_s^{-1}q^{\rho\sigma}\pd_{\rho}h_0\pd_{\sigma}k_0
-\frac{1}{2}\gamma_{ij}\alpha\left[\lap_{\Xsp}h_0
+\left(h_rh_s\right)^{-1}\lap_{\Zsp}h_1\right]\nn\\
&&~~~~-\frac{1}{2}\gamma_{ij}(\beta-1)h_rh_s^{-1}
\left[\lap_{\Xsp}k_0+\left(h_rh_s\right)^{-1}\lap_{\Zsp}k_1\right]=0,
 \label{1:c2Einstein-ij:Eq}\\
&&R_{mn}(\Ysp_2)-\frac{1}{2}\left(\alpha+\beta-1\right)
w_{mn}h_r^{-1}q^{\rho\sigma}\pd_{\rho}h_0\pd_{\sigma}k_0
-\frac{1}{2}\beta w_{mn}
\left[\lap_{\Xsp}k_0+\left(h_rh_s\right)^{-1}\lap_{\Zsp}k_1\right]\nn\\
&&~~~~-\frac{1}{2}(\alpha-1)w_{mn}h_r^{-1}h_s
\left[\lap_{\Xsp}h_0+\left(h_rh_s\right)^{-1}\lap_{\Zsp}h_1\right]=0,
 \label{1:c2Einstein-mn:Eq}\\
&&R_{ab}(\Zsp)-\frac{1}{2}\left(\alpha+\beta\right)
u_{ab}q^{\rho\sigma}\pd_{\rho}h_0\pd_{\sigma}k_0
-\frac{1}{2}\alpha u_{ab}h_s
\left[\lap_{\Xsp}h_0+\left(h_rh_s\right)^{-1}\lap_{\Zsp}h_1\right]\nn\\
&&~~~~-\frac{1}{2}\beta u_{ab}h_r\left[\lap_{\Xsp}k_0
+\left(h_rh_s\right)^{-1}\lap_{\Zsp}k_1\right]=0.
  \label{1:c2Einstein-ab:Eq}
}

Finally, we should consider the scalar field equation.
Substituting Eqs.~(\ref{1:ansatz:Eq}), (\ref{1:warp:Eq}) and
the intersection rule $\chi=0$ into Eq.~(\ref{D:scalar:Eq}), we obtain
\Eqr{
&&h_r^{-\alpha}h_s^{-\beta}\left[\epsilon_rc_r\left\{h_s\triangle_{\Xsp}h_0
+q^{\rho\sigma}\pd_{\rho}h_0\pd_{\sigma}k_0
+h_r^{-1}\triangle_{\Zsp}h_1\right\}\right.\nn\\
&&\left.~~~~+\epsilon_sc_s
\left\{h_r\triangle_{\Xsp}k_0+q^{\rho\sigma}\pd_{\rho}h_0\pd_{\sigma}k_0
+h_s^{-1}\lap_{\Zsp}k_1\right\}
\right]=0.
  \label{1:scalar equation:Eq}
}
Thus, the warp factors $h_r$ and $h_s$ should satisfy the equations
\Eqrsubl{1:scalar solution:Eq}{
&&\triangle_{\Xsp}h_0=0,~~~ \triangle_{\Zsp}h_1=0,~~~
\triangle_{\Zsp}h_s=0,~~~{\rm for}~~\pd_{\mu}k_0=0,
   \label{1:scalar solution1:Eq}\\
&&\triangle_{\Xsp}k_0=0,~~~ \triangle_{\Zsp}k_1=0,~~~
\triangle_{\Zsp}h_r=0,~~~{\rm for}~~\pd_{\mu}h_0=0.
   \label{1:scalar solution2:Eq}
}

Combining these, we find that these field equations lead to
\cite{Maeda:2009zi}
\Eqrsubl{1:solution of Einstein:Eq}{
&&R_{\mu\nu}(\Xsp)=0,~~~~R_{ij}(\Ysp_1)=0,~~~~
R_{mn}(\Ysp_2)=0,~~~~R_{ab}(\Zsp)=0,
   \label{1:Ricci:Eq}\\
&&h_r=h_0(x)+h_1(z),~~~~h_s=k_0(x)+k_1(z),
   \label{1:h:Eq}\\
&&D_{\mu}D_{\nu}h_0=0,~~~\triangle_{\Zsp}h_1=0,~~~ \triangle_{\Zsp}h_s=0,
~~~~{\rm for}~~\pd_{\mu}h_s=0,
   \label{1:warp2:Eq}\\
&&D_{\mu}D_{\nu}k_0=0,~~~\triangle_{\Zsp}k_1=0,~~~ \triangle_{\Zsp}h_r=0,
~~~~{\rm for}~~\pd_{\mu}h_r=0.
   \label{1:warp3:Eq}
 }
If $F_{\left(p_r+2\right)}=0$ and $F_{\left(p_s+2\right)}=0$,
the functions $h_1$ and $k_1$ become trivial.
Namely, the $D$-dimensional spacetime is no longer warped~
\cite{Kodama:2005fz, Kodama:2005cz}.

As a special example, let us consider the case
\Eq{
q_{\mu\nu}=\eta_{\mu\nu}\,,~~~\gamma_{ij}=\delta_{ij}\,,~~~
w_{mn}=\delta_{mn}\,,~~~u_{ab}=\delta_{ab}\,,
 \label{1:flat metric:Eq}
 }
where $\eta_{\mu\nu}$ is the $(p+1)$-dimensional
Minkowski metric and $\delta_{ij}$, $\delta_{mn}$, $\delta_{ab}$ are
the $(p_s-p)$-, $(p_r-p)$-, and $(D+p-p_r-p_s-1)$-dimensional Euclidean metrics,
respectively. For $\pd_{\mu}h_s=0$, the solution for $h_r$ and $h_s$ can be
obtained explicitly as
\Eqrsubl{1:solution:Eq}{
h_r(x, z)&=&A_{\mu}x^{\mu}+B
+\sum_{l}\frac{M_l}{|\bm z-\bm z_l|^{D+p-p_r-p_s-3}},
 \label{1:solution-r:Eq}\\
h_s(z)&=&C+\sum_{c}\frac{M_c}{|\bm z-\bm z_c|^{D+p-p_r-p_s-3}},
 \label{1:solution-s:Eq}
}
where $A_{\mu}$, $B$, $C$, $M_l$, and $M_c$ are constant parameters,
and $\bm z_l$ and $\bm z_c$ are constant vectors representing the positions of the branes.
We can also choose the solution in which the $p_s$-brane part
depends on both $t$ and $z$. Then, we have
\Eqrsubl{1:solution2:Eq}{
h_r(z)&=&B
+\sum_{l}\frac{M_l}{|\bm z-\bm z_l|^{D+p-p_r-p_s-3}},
 \label{1:solution2-r:Eq}\\
h_s(x,z)&=&A_{\mu}x^{\mu}+C+\sum_{c}\frac{M_c}{|\bm z-\bm z_c|^{D+p-p_r-p_s-3}}.
 \label{1:solution2-s:Eq}
}

Now we discuss the intersecting brane solutions in eleven-dimensional
supergravity and in ten-dimensional string theories.
For the M-branes in eleven-dimensional supergravity, there is
4-form field strength without dilaton, so the intersection rule $\chi=0$ gives
\Eq{
p=\frac{(p_r+1)(p_s+1)}{9}-1,
    \label{1:chi2:Eq}
}
where $p$ denotes the number of overlapping dimensions of the
$p_r$ and $p_s$ branes. Then we get the intersections involving the
M2 and M5-branes \cite{Argurio:1997gt,Ohta:1997gw}
\Eq{
{\rm M2}\cap {\rm M2}=0,~~~~{\rm M2}\cap {\rm M5}=1,~~~~
{\rm M5}\cap {\rm M5}=3.~~~~
   \label{1:int M:Eq}
}

For the ten-dimensional string theories, the couplings to dilaton
for the RR-charged D-branes are given by
\Eq{
\epsilon_rc_r=\frac{1}{2}\left(3-p_r\right),~~~~
\epsilon_sc_s=\frac{1}{2}\left(3-p_s\right).
}
The condition $\chi=0$ then gives
\Eq{
p=\frac{1}{2}\left(p_r+p_s-4\right).
}
The intersections for the D-branes are thus given by~\cite{Argurio:1997gt,Ohta:1997gw}
\Eq{
%{\rm D}\bar{p}\cap {\rm D}\bar{p}' = \frac{1}{2}(\bar{p}+\bar{p}')-2.
{\rm D}p_r\cap {\rm D}p_s = \frac{1}{2}(p_r+p_s)-2.
   \label{1:int D:Eq}
}
We finally consider the intersections for NS-branes. The parameters $c_r$ for
fundamental string (F1) and solitonic 5-brane are $\epsilon_1c_1=-1$
(for F1) and $\epsilon_5c_5=1$ (for NS5), respectively.
Then the intersections involving the
F1 and NS5-branes are \cite{Argurio:1997gt,Ohta:1997gw}
\Eqrsubl{1:int DNS:Eq}{
&&
%{\rm F1}\cap {\rm F1}=-1,~~~~
{\rm F1}\cap {\rm NS5}=1,~~~~
{\rm NS5}\cap {\rm NS5}=3,
   \label{1:int NS:Eq}\\
&&{\rm F1}\cap {\rm D}\bar{p}=0,\\
&&{\rm D}\bar{p}\cap {\rm NS}5=\bar{p}-1,~~~~1\le\bar{p}\le 6.
}
There is no solution for the F1-F1 and D0-NS5 intersecting brane systems
because the numbers of space dimensions for
each pairwise overlap are negative by the intersection rule.
%%%%%%%%%%%%%%%%%%%%%
%%%%%%%%%%%%%%%%%%%%%
%%%%%%%%%%%%%%%%%%%%%
%T2> CaseII
\subsection{Case (II)}
  \label{sub:II}

We next consider case (II).
For this class, the $D$-dimensional metric ansatz (\ref{D:metric:Eq}) gives
\Eqr{
ds^2&=&h^{\alpha}_r(x, y)h_s^{\beta}(x, z)\left[
h_r^{-1}(x, y)h_s^{-1}(x, z)q_{\mu\nu}(\Xsp)dx^{\mu}dx^{\nu}
  +h_s^{-1}(x, z)\gamma_{ij}(\Ysp_1)dy^idy^j\right.\nn\\
&&\left. +h_r^{-1}(x, y)w_{mn}(\Ysp_2)dv^{m}dv^{n}
  +u_{ab}(\Zsp)dz^adz^b\right].
 \label{2:metric:Eq}
}
We also take the following ansatz for the scalar field $\phi$ and
the gauge field strengths:
\Eqrsubl{2:ansatz:Eq}{
\e^{\phi}&=&h_r^{\epsilon_rc_r/2}\,
h_s^{\epsilon_sc_s/2},
  \label{2:ansatz for scalar:Eq}\\
F_{\left(p_r+2\right)}&=&d\left[h^{-1}_r(x, y)\right]\wedge\Omega(\Xsp)
\wedge\Omega(\Ysp_2),
  \label{2:ansatz for gauge-r:Eq}\\
F_{\left(p_s+2\right)}&=&d\left[h^{-1}_s(x, z)\right]\wedge\Omega(\Xsp)
\wedge\Omega(\Ysp_1),
  \label{2:ansatz for gauge-s:Eq}
}
where $\Omega(\Xsp)$, $\Omega(\Ysp_1)$, and $\Omega(\Ysp_2)$
are defined in (\ref{1:volume:Eq}).
Since we use the same procedure as in Sec.~\ref{sub:I}, we can derive
the intersection rule $\chi=0$ from the field equations.
For $\chi=0$, it is easy to show that the field equations reduce to
\Eqrsubl{2:solution of Einstein:Eq}{
&&R_{\mu\nu}(\Xsp)=0,~~~~R_{ij}(\Ysp_1)=0,~~~~
R_{mn}(\Ysp_2)=0,~~~~R_{ab}(\Zsp)=0,
   \label{2:Ricci:Eq}\\
&&h_r=h_0(x)+h_1(y),~~~~h_s=h_s(z),
   \label{2:h:Eq}\\
&&D_{\mu}D_{\nu}h_0=0,~~~\triangle_{\Ysp_1}h_1=0,~~~ \triangle_{\Zsp}h_s=0,
~~~~\pd_{\mu}h_s=0,
   \label{2:warp2:Eq}
 }
where $\triangle_{\Ysp_1}$ is the Laplace
operators on the space of $\Ysp_1$.
If $F_{\left(p_r+2\right)}\ne 0$ and $F_{\left(p_s+2\right)}\ne 0$,
the functions $h_1$ and $k_1$ are nontrivial.

Let us consider the following case in more detail:
\Eq{
q_{\mu\nu}=\eta_{\mu\nu}\,,~~~\gamma_{ij}=\delta_{ij}\,,~~~
w_{mn}=\delta_{mn}\,,~~~u_{ab}=\delta_{ab}\,,
 \label{2:flat metric:Eq}
 }
where $\eta_{\mu\nu}$ is the $(p+1)$-dimensional
Minkowski metric and $\delta_{ij}$, $\delta_{mn}$, $\delta_{ab}$ are
the $(p_s-p)$-, $(p_r-p)$-, and $(D+p-p_r-p_s-1)$-dimensional Euclidean metrics,
respectively. For $\pd_{\mu}h_s=0$, the solution for $h_r$ and
$h_s$ can be obtained explicitly as
\Eqrsubl{2:solution:Eq}{
h_r(x, y)&=&A_{\mu}x^{\mu}+B
+\sum_{l}\frac{M_l}{|\bm y-\bm y_l|^{p_s-p-2}},
 \label{2:solution-r:Eq}\\
h_s(z)&=&C+\sum_{c}\frac{M_c}{|\bm z-\bm z_c|^{D+p-p_r-p_s-3}},
 \label{2:solution-s:Eq}
}
where $A_{\mu}$, $B$, $C$, $\bm y_l$, $\bm z_c$, $M_l$, 
and $M_c$ are constant parameters.

One can easily get the solution
for $\pd_{\mu}h_r=0$ and $\pd_{\mu}h_s\ne 0$
if the roles of $\Ysp_1$ and $\Ysp_2$ are exchanged.
The solution of field equations is thus expressed as
\Eqrsubl{2:solution2:Eq}{
h_r(z)&=&B
+\sum_{l}
       \frac{M_l}{|\bm z-\bm z_l|^{D+p-p_s-p_r-3}},
 \label{2:solution2-r:Eq}\\
h_s(x,v)&=&A_{\mu}x^{\mu}+C+\sum_{c}\frac{M_c}{|\bm v-\bm v_c|^{p_r-p-2}}.
 \label{2:solution2-s:Eq}
}

Since the dynamical solution (\ref{2:solution:Eq})
obeys the same intersection rule $\chi=0$,
the intersections of M-branes in eleven-dimensional
supergravity and D-branes in ten-dimensional string theories
are given as (\ref{1:int M:Eq}), (\ref{1:int D:Eq}), and (\ref{1:int DNS:Eq}).

%%%%%%%%%%%%%%%%%%%%%%%%
%%%%%%%%%%%%%%%%%%%%%%%%
%%%%%%%%%%%%%%%%%%%%%%%%
%T2> CaseIII
\subsection{Case (III)}
  \label{sub:III}

Finally, we consider case (III).
For this class, the $D$-dimensional metric ansatz~(\ref{D:metric:Eq}) reduces to
\Eqr{
ds^2&=&h^{\alpha}_r(x, y)h_s^{\beta}(x, v)\left[
h_r^{-1}(x, y)h_s^{-1}(x, v)q_{\mu\nu}(\Xsp)dx^{\mu}dx^{\nu}
  +h_s^{-1}(x, v)\gamma_{ij}(\Ysp_1)dy^idy^j\right.\nn\\
&&\left. +h_r^{-1}(x, y)w_{mn}(\Ysp_2)dv^{m}dv^{n}
  +u_{ab}(\Zsp)dz^adz^b\right].
 \label{3:metric:Eq}
}
We also assume that the scalar field $\phi$ and
the gauge field strengths are given as
\Eqrsubl{3:ansatz:Eq}{
\e^{\phi}&=&h_r^{\epsilon_rc_r/2}\,
h_s^{\epsilon_sc_s/2},
  \label{3:ansatz for scalar:Eq}\\
F_{\left(p_r+2\right)}&=&h_s\,d\left[h^{-1}_r(x, y)\right]\wedge\Omega(\Xsp)
\wedge\Omega(\Ysp_2),
  \label{3:ansatz for gauge-r:Eq}\\
F_{\left(p_s+2\right)}&=&h_r\,d\left[h^{-1}_s(x, v)\right]\wedge\Omega(\Xsp)
\wedge\Omega(\Ysp_1),
  \label{3:ansatz for gauge-s:Eq}
}
where $\Omega(\Xsp)$, $\Omega(\Ysp_1)$, and $\Omega(\Ysp_2)$
denote the volume $(p+1)$-, $(p_s-p)$-, and $(p_r-p)$-forms, respectively.

Under the assumption, the field equations give the intersection rule $\chi=-2$.
This is different from the usual rule applicable to the cases (I) and (II).
Upon using the intersection rule $\chi=-2$, it is easy to show that the field
equations reduce to
\Eqrsubl{3:solution of Einstein:Eq}{
&&R_{\mu\nu}(\Xsp)=0,~~~~R_{ij}(\Ysp_1)=0,~~~~
R_{mn}(\Ysp_2)=0,~~~~R_{ab}(\Zsp)=0,
   \label{3:Ricci:Eq}\\
&&h_r=h_0(x)+h_1(y),~~~~h_s=k_0(x)+k_1(v),
   \label{3:h:Eq}\\
&&D_{\mu}D_{\nu}h_0=0,~~~\triangle_{\Ysp_1}h_1=0,~~~ \triangle_{\Ysp_2}h_s=0,
~~~~{\rm for}~~\pd_{\mu}h_s=0,
   \label{3:warp2:Eq}\\
&&D_{\mu}D_{\nu}k_0=0,~~~\triangle_{\Ysp_1}h_r=0,~~~ \triangle_{\Ysp_2}k_1=0,
~~~~{\rm for}~~\pd_{\mu}h_r=0.
   \label{3:warp3:Eq}
 }
where $\triangle_{\Ysp_1}$ and $\triangle_{\Ysp_2}$ are the Laplace
operators on the spaces of $\Ysp_1$ and $\Ysp_2$, respectively.
The functions $h_1$ and $k_1$ are nontrivial for
$F_{\left(p_r+2\right)}\ne 0$ and $F_{\left(p_s+2\right)}\ne 0$.

Now we consider the case
\Eq{
q_{\mu\nu}=\eta_{\mu\nu}\,,~~~\gamma_{ij}=\delta_{ij}\,,~~~
w_{mn}=\delta_{mn}\,,~~~u_{ab}=\delta_{ab}\,,
 \label{3:flat metric:Eq}
 }
where $\eta_{\mu\nu}$ is the $(p+1)$-dimensional
Minkowski metric and $\delta_{ij}$, $\delta_{mn}$, $\delta_{ab}$ are
the $(p_s-p)$-, $(p_r-p)$-, and $(D+p-p_r-p_s-1)$-dimensional Euclidean metrics,
respectively.
For $\pd_{\mu}h_s=0$, the solution for $h_r$ and $h_s$ can be obtained
explicitly as
\Eqrsubl{3:solution:Eq}{
h_r(x, y)&=&A_{\mu}x^{\mu}+B
+\sum_{l}\frac{M_l}{|\bm y-\bm y_l|^{p_s-p-2}},
 \label{3:solution-r:Eq}\\
h_s(v)&=&C+\sum_{c}\frac{M_c}{|\bm v-\bm v_c|^{p_r-p-2}},
 \label{3:solution-s:Eq}
}
where $A_{\mu}$, $B$, $C$, $\bm y_l$, $\bm v_c$,
$M_l$, and $M_c$ are constant parameters.
We can also get the solution in which the function $h_s$
depends on both $t$ and $v$. The solution
(\ref{3:solution:Eq}) is replaced by
\Eqrsubl{3:solution2:Eq}{
h_r(y)&=&B
+\sum_{l}\frac{M_l}{|\bm y-\bm y_l|^{p_s-p-2}},
 \label{3:solution2-r:Eq}\\
h_s(x,v)&=&A_{\mu}x^{\mu}+C+\sum_{c}\frac{M_c}{|\bm v-\bm v_c|^{p_r-p-2}}.
 \label{3:solution2-s:Eq}
}

Let us consider the intersecting brane solutions in eleven-dimensional
supergravity and in ten-dimensional string theories.
We first discuss the intersections of M-branes in eleven-dimensional
supergravity. The intersection rule $\chi=-2$ leads to 
\cite{Gauntlett:1996pb, Edelstein:1998vs, Edelstein:2004tp}
\Eq{
p=\frac{(p_r+1)(p_s+1)}{9}-3.
    \label{3:chi2:Eq}
}
Then we get the intersection involving the M5-brane
\Eq{
%{\rm M2}\cap {\rm M2}=-2,~~~~
%{\rm M2}\cap {\rm M5}=-1,~~~~
{\rm M5}\cap {\rm M5}=1.~~~~
   \label{3:int M:Eq}
}
%There are no intersections for M2-M2 and M2-M5 branes.
%In particular, this rule
Equation~\eqref{3:chi2:Eq} tells us that the numbers of
intersection for M2-M2 and M2-M5 branes are negative,
which means that there is no intersecting solution of these brane systems.

Next we consider the intersection in the ten-dimensional string theory.
The couplings to dilaton for the RR-charged D-branes are
\Eq{
\epsilon_rc_r=\frac{1}{2}\left(3-p_r\right),~~~~
\epsilon_sc_s=\frac{1}{2}\left(3-p_s\right),
}
and the condition $\chi=-2$ is expressed as
\Eq{
p=\frac{1}{2}\left(p_r+p_s-8\right).
}
The intersections for the RR-charged D-branes are thus given by
\Eq{
%{\rm D}\bar{p}\cap {\rm D}\bar{p}' =\frac{1}{2}(\bar{p}+\bar{p}')-4.
{\rm D}p_r\cap {\rm D}p_s = \frac{1}{2}(p_r+p_s)-4.
  \label{3:int D:Eq}
}
We finally consider the intersections for NS-branes. The parameters $c_r$ for
fundamental string (F1) and solitonic 5-brane are $\epsilon_1c_1=-1$
for F1 and $\epsilon_5c_5=1$ for NS5, respectively.
Then the intersection with F1-brane is forbidden by the intersection rule.
The intersections involving the NS5-branes are
\Eqrsubl{3:int DNS:Eq}{
&&
%{\rm F1}\cap {\rm NS5}=-1,~~~~
{\rm NS5}\cap {\rm NS5}=1,
   \label{3:int NS:Eq}\\
%&&{\rm F1}\cap {\rm D}\bar{p}=-2,\\
&&{\rm D}\bar{p}\cap {\rm NS}5=\bar{p}-3,~~~~3\le \bar{p}\le 8.
}
There is no brane solution involving other intersections because
the numbers of space dimensions for
each pairwise overlap become negative by the intersection rule.

%======================================%
%<<<<<<<<<<<<< SECTION 3 >>>>>>>>>>>>>>%
%======================================%
%T1>Cosmology
\section{Cosmology}
\label{sec:cosmology}

In this section, we discuss the application of the above solutions to
four-dimensional cosmology.
We assume an isotropic and homogeneous three-space in the four-dimensional
spacetime known as Friedmann-Lema\^itre-Robertson-Walker (FLRW) universe,
and do not discuss solutions which break these properties after compactification.
In what follows, we concentrate on the $(p+1)$-dimensional Minkowski spacetime
with $q_{\mu\nu}(\Xsp)=\eta_{\mu\nu}(\Xsp)$, and
drop the coordinate dependence on $\Xsp$ space except for the time.

The $D$-dimensional metric (\ref{D:metric:Eq}) can be expressed as
\Eq{
ds^2=-hdt^2+ds^2(\tilde{\Xsp})+ds^2(\Ysp_1)+
ds^2(\Ysp_2)+ds^2(\Zsp),
   \label{b:metric:Eq}
}
where we have defined
\Eqrsubl{b:metric1:Eq}{
ds^2(\tilde{\Xsp})&\equiv&h\delta_{PQ}(\tilde{\Xsp})d\theta^{P}d\theta^{Q},\\
ds^2(\Ysp_1)&\equiv&h^{\alpha}_rh_s^{\beta-1}\gamma_{ij}(\Ysp_1)dy^idy^j,\\
ds^2(\Ysp_2)&\equiv&h^{\alpha-1}_rh_s^{\beta}w_{mn}(\Ysp_2)dv^{m}dv^{n},\\
ds^2(\Zsp)&\equiv&h^{\alpha}_rh_s^{\beta}u_{ab}(\Zsp)dz^adz^b,\\
h&\equiv&h^{\alpha-1}_rh_s^{\beta-1}.
 }
Here $\delta_{PQ}(\tilde{\Xsp})$ is the $p$-dimensional Euclidean
metric, and $\theta^P$ denotes the coordinate of the $p$-dimensional Euclid
space $\tilde{\Xsp}$. In the following, we assume $\pd_{\mu}h_s=0$ and
set $h_r=At+h_1$. The $D$-dimensional metric
(\ref{b:metric1:Eq}) can be written as
\Eqr{
ds^2&=&h_s^{\beta-1}\left[1+\left(\frac{\tau}{\tau_0}\right)^{-2/(\alpha+1)}
h_1\right]^{\alpha-1}
\left[-d\tau^2+\left(\frac{\tau}{\tau_0}\right)^{2(\alpha-1)/(\alpha+1)}
\delta_{PQ}(\tilde{\Xsp})d\theta^Pd\theta^Q\right.\nn\\
&&+\left\{1+\left(\frac{\tau}{\tau_0}\right)^{-2/(\alpha+1)}h_1\right\}
\left(\frac{\tau}{\tau_0}\right)^{2\alpha/(\alpha+1)}
\gamma_{ij}(\Ysp_1)dy^idy^j\nn\\
&&+h_s\left(\frac{\tau}{\tau_0}\right)^{2(\alpha-1)/(\alpha+1)}
w_{mn}(\Ysp_2)dv^mdv^n\nn\\
&&\left.+h_s\left\{1+\left(\frac{\tau}{\tau_0}\right)^{-2/(\alpha+1)}
h_1\right\}\left(\frac{\tau}{\tau_0}\right)^{2\alpha/(\alpha+1)}
u_{ab}(\Zsp)dz^adz^b\right]
   \label{b:metric-a:Eq}
}
where we have introduced the cosmic time $\tau$ defined by
\Eq{
\frac{\tau}{\tau_0}=\left(At\right)^{(\alpha+1)/2},~~~~\tau_0=
\frac{2}{\left(\alpha+1\right)A}.
}
On the other hand, for $h_s(t, v)=At+k_1(v)$, the metric
(\ref{b:metric1:Eq}) is given by replacing $\alpha$ and $h_1$ with $\beta$
and $k_1$.

Now we apply these solutions to lower-dimensional effective theory.
We compactify $d(\equiv d_1+d_2+d_3+d_4)$ dimensions to fit our universe,
where $d_1$, $d_2$, $d_3$, and $d_4$ denote the compactified dimensions with
respect to the $\tilde{\Xsp}$, $\Ysp_1$, $\Ysp_2$, and $\Zsp$ spaces.
The metric (\ref{b:metric:Eq}) is then described by
\Eq{
ds^2=ds^2(\Msp)+ds^2(\Nsp),
   \label{b:metric2:Eq}
}
where $ds^2(\Msp)$ is the $(D-d)$-dimensional metric and
$ds^2(\Nsp)$ is the metric of compactified dimensions.

By the conformal transformation
\Eq{
ds^2(\Msp)=h_r^Bh_s^Cds^2(\bar{\Msp}),
}
we can rewrite the $(D-d)$-dimensional metric in the Einstein frame.
Here $B$ and $C$ are
\Eq{
B=\frac{-\alpha d+d_1+d_3}{D-d-2},~~~~~~C=\frac{-\beta d+d_2+d_4}{D-d-2}.
}
Hence, the $(D-d)$-dimensional metric in the Einstein frame is
\Eqr{
ds^2(\bar{\Msp})&=&h_r^{B'}h_s^{C'}\left[-d\tau^2+\delta_{P'Q'}
(\tilde{\Xsp}')d\theta^{P'}d\theta^{Q'}
+h_r\gamma_{k'l'}({\Ysp_1}')dy^{k'}dy^{l'}\right.\nn\\
&&\left.+h_sw_{m'n'}({\Ysp_2}')dv^{m'}dv^{n'}
+h_rh_su_{a'b'}({\Zsp}')dz^{a'}dz^{b'}\right],
  \label{b:metric-m:Eq}
}
where $B'$ and $C'$ are defined by $B'=-B+\alpha-1$ and $C'=-C+\beta-1$,
and $\tilde{\Xsp}'$, ${\Ysp_1}'$, ${\Ysp_2}'$, and ${\Zsp}'$ denote
the $(p-d_1)$-, $(p_s-p-d_2)$-, $(p_r-p-d_3)$-, and
$(D+p-p_r-p_s-d_4)$-dimensional spaces, respectively.

For $h_r=At+h_1$, the metric (\ref{b:metric-m:Eq}) is thus rewritten as
\Eqr{
ds^2(\bar{\Msp})&=&h_s^{C'}\left[1+\left(\frac{\tau}{\tau_0}\right)
^{-2/(B'+2)}h_1\right]^{B'}\left[-d\tau^2+
\left(\frac{\tau}{\tau_0}\right)^{2B'/(B'+2)}
\delta_{P'Q'}(\tilde{\Xsp}')d\theta^{P'}d\theta^{Q'}\right.\nn\\
&&+\left\{1+\left(\frac{\tau}{\tau_0}\right)^{-2/(B'+2)}h_1\right\}
\left(\frac{\tau}{\tau_0}\right)^{2(B'+1)/(B'+2)}
\gamma_{k'l'}({\Ysp_1}')dy^{k'}dy^{l'}\nn\\
&&+h_s\left(\frac{\tau}{\tau_0}\right)^{2B'/(B'+2)}
w_{m'n'}({\Ysp_2}')dv^{m'}dv^{n'}\nn\\
&&\left.+h_s\left\{1+\left(\frac{\tau}{\tau_0}\right)^{-2/(B'+2)}
h_1\right\}\left(\frac{\tau}{\tau_0}\right)^{2(B'+1)/(B'+2)}
u_{a'b'}({\Zsp}')dz^{a'}dz^{b'}\right],
  \label{b:metric-mr:Eq}
}
where the cosmic time $\tau$ is defined by
\Eq{
\frac{\tau}{\tau_0}=\left(At\right)^{(B'+2)/2},~~~~\tau_0=
\frac{2}{\left(B'+2\right)A}.
}
For $h_s=At+k_1$ and $\pd_{\mu}h_r=0$,
we can also get results similar to
(\ref{b:metric-a:Eq}) and (\ref{b:metric-mr:Eq}).

We list the FLRW cosmological solutions with an isotropic and homogeneous
three-space for the solutions~(\ref{b:metric-mr:Eq})
in Table \ref{table_1} for M-branes,
Tables~\ref{table_2}-\ref{table_9} for D-branes,
and \ref{table_10}-\ref{table_14} for F1 and NS5-branes.
The power exponents of the scale factor of possible four-dimensional cosmological
models are given by $a(\tilde{\Msp})\propto \tau^{\lambda(\tilde{\Msp})}$,
where $\tau$ is the cosmic time, and $a(\tilde{\Msp})$ and
$a_{\rm E}(\tilde{\Msp})$ denote the scale factors of the space
$\tilde{\Msp}$ in Jordan and Einstein frames with the exponents carrying
the same suffices, respectively.
Here $\tilde{\Msp}$ denotes the spatial part of spacetime $\Msp$
and includes our four-dimensional universe besides the time coordinate.
The mark $\surd$ in the tables shows which brane is time dependent.

Since the time dependence in the metric comes from only one
brane in the intersections, the obtained expansion law is simple.
In order to find an expanding universe, one may have to
compactify the vacuum bulk space as well as the brane worldvolume.
Unfortunately, we find that the fastest expanding case in the Jordan frame
has the power $a\propto \tau^{7/15}$, which is too small to give a realistic
expansion law like that in the matter dominated era ($a\propto \tau^{2/3}$)
or that in the radiation dominated era ($a\propto \tau^{1/2}$).
Note that all these cases correspond to the solutions involving D6-branes,
given in Tables~\ref{table_4}-\ref{table_9} and \ref{table_11}-\ref{table_13}.

When we compactify the extra dimensions and go to the four-dimensional
Einstein frame, the power exponents are different depending on how we compactify
the extra dimensions even within one solution.
We give the power exponent of the fastest expansion of our four-dimensional universe
in the Einstein frame in Tables~\ref{table_14}-\ref{table_25}.
We again see that the expansion is too small.
Hence, we have to conclude that in order to find a realistic expansion
of the universe in this type of models, one has to include
additional ``matter" fields on the brane.

%======================================%
%<<<<<<<<<<<<< SECTION 4 >>>>>>>>>>>>>>%
%======================================%

%T1>Brane world approach

\section{Brane world approach}
\label{sec:app-to-bw}

\subsection{Construction of the brane world model}

In this section, we discuss the applications
of our solutions to construct the brane world models.
Starting from a given ten- or eleven-dimensional solution,
we compactify the trivial extra dimensions.
After the reduction, we further move to the Einstein frame.
Then, there are the ordinary four-dimensional spacetime
and extra dimensions.
The following procedure depends on the number of codimensions $n$.

For $n>2$, the brane is so singular that one cannot put the ordinary matter
and
we employ the cut-copy-paste method as a way of regularization,
which is explained later. %(see Fig. \ref{BW} for an image).
As a result, the original brane with $n$ codimensions
is replaced with a codimension-one
object including the internal angular dimensions.
For $n=2$, there is a curvature singularity at the infinity
due to the logarithmic spatial dependence of the metric
and we do not discuss this case any more.
For $n=1$, the spacetime is regular.
In all cases, due to the presence of the time dependence,
there can be the other singularity in the future or past,
where $h_r=0$ in brane solutions.
For the moment, we focus on the prescription for the brane and
will discuss the behavior of the brane world near the
time dependent singularity
in the next subsection.
In both cases,
the boundary of the bulk spacetime
is a codimension-one hypersurface,
which is called ``brane world'' in the rest.
By construction, the extra space is $Z_2$-symmetric
with respect to the brane world.
%%%%%%%%%%%%%%%%%%%%%%%%%%%%%%%%%%%%%%%%%%%%%%%%%%%%
%\begin{figure}[tbh]
%\begin{center}
%\includegraphics[width=7cm,clip]{bw1.eps}
%\hskip 1cm
%\includegraphics[width=7cm,clip]{bw2.eps}
%\\
%{\LARGE (a)} \hskip 6.5cm
%{\LARGE (b)}
%\\
%\includegraphics[width=9cm,clip]{bw3.eps}\\
%{\LARGE (c)}
%\caption{\baselineskip 12pt
%A broad image of our cut-copy-paste procedure to construct the brane world when%$n>2$.
%The radial direction is denoted by $\xi$,
%while the circular direction denotes the angular dimensions
%collectively.
%Each figure shows\\
%(a) The original solution including a brane with higher codimensions. \\
%(b) Removing the piece containing brane at $\xi=\xi_0>0$.\\
%(c) Gluing the remaining piece to its identical copy.
%The brane worldvolume (the dashed curve) contains the internal angular dimensio%ns.
%}
%\label{BW}
%\end{center}
%\vspace{-5mm}
%\end{figure}
%%%%%%%%%%%%%%%%%%%%%%%%%%%%%%%%%%%%%%%%%%%%%%%%%%%

We consider the time dependent Einstein-frame metric
with $n$ conformally flat extra dimensions,
\begin{eqnarray}
ds^2=-d(t,\xi)^2 dt^2
     +a(t,\xi)^2 \delta_{ij}d{\cal X}^i d{\cal X}^j
     +f(t, \xi)^2\Big(d\xi^2
     +\xi^2{\cal G}_{ab}d\theta^a d\theta^b\Big)\,,
\label{metric_ge}
\end{eqnarray}
where metric ${\cal G}_{ab}$ [$a=1,2,\cdots,(n-1)$]
represents the unit $(n-1)$ sphere.
The coordinates ($\xi$, $\theta^a$) and
${\cal X}^i$ denote the radial and angular directions of the extra space,
and the ordinary three-space, respectively.
We assume that in the original intersecting brane solution,
the brane is at $\xi=0$.
For $n>2$, we need a regularization to put the matter on the brane
and we employ the cut-copy-paste method:
%to construct the brane world $\xi=\xi_0>0$.% (see Fig. \ref{BW}).
\begin{description}

\item{(a) The region including the brane $0\leq \xi<\xi_0$ is removed.}

\item{(b) The remaining piece is glued to its identical copy at $\xi=\xi_0$.}

\end{description}
In this way, the brane is replaced with a codimension-one
brane world including the extra angular dimensions.
For $n=1$, we assume that the brane world is at the place where
the original brane source exists.

The brane world moves along the trajectory $(t(\tau),\xi(\tau))$,
where $\tau$ satisfies
\begin{eqnarray}
-d(t(\tau), \xi(\tau))^2
\dot{t}{}^2
+
f(t(\tau), \xi(\tau))^2\dot{\xi}{}^2=-1.
\label{normalization}
\end{eqnarray}
The ``dot'' denotes the derivative with respect to $\tau$.
Thus, the induced metric becomes cosmological
\begin{eqnarray}
ds_{\rm ind}^2
=- d\tau^2+a(t(\tau),\xi(\tau))^2 \delta_{ij}d{\cal X}^i d{\cal X}^j
  +\xi(\tau)^2f(t(\tau), \xi(\tau))^2{\cal G}_{ab}d\theta^a d\theta^b\,.
\end{eqnarray}
%{\bf
In the case of $n>2$,
since ${\cal G}_{ab}$ denotes the $(n-1)$-dimensional sphere,
the corresponding dimensions in the brane worldvolume are automatically
compact as long as the scale factor in these directions $\xi f$ is finite.
%}
To construct the 
brane world by our cut-copy-paste method, further simplifications
should be required:
We first assume that each function
$h_r$ and $h_s$ in Eqs.~(\ref{1:solution:Eq}), (\ref{2:solution:Eq})
and (\ref{3:solution:Eq}) contains only a contribution from a single brane,
with charge $M_r$ and $M_s$, respectively.
We then impose additional restrictions in each of the cases (I)-(III).
\begin{enumerate}

\item{Case (I)}

Both functions $h_r$ and $h_s$ depend on  $\vect{z}$:
\bea
h_r(\vect{z},t)=A_0 t+B
              +\frac{M_r}{| \vect{z}-\vect{z}_r|^{n-2}},\quad
%and
h_s(\vect{z})=C
              +\frac{M_s}{| \vect{z}-\vect{z}_s|^{n-2}},
\eea
(see Eq. (\ref{1:solution:Eq})).
The positions of branes, $\vect{z}_{r}$ and $\vect{z}_s$,
are different in general. Our cut-copy-paste procedure works
when the extra space is spherically symmetric.
To realize the spherical symmetry, we put the branes at the same
place $\vect{z}_r=\vect{z}_s=0$ and then set $\xi=|\vect{z}|$.
Alternatively, we may choose $\vect{z}_s=0$ for $M_r=0$
and $\vect{z}_r=0$ for $M_s=0$, with $\xi=|\vect{z}|$,
but we focus on the former general case with $M_s M_r \neq 0$ here.
For $n=1$, we may take $z_r\neq z_s$.

\item{Case (II)}

Here $h_r$ and $h_s$ depend on $\,\vect{y}$ and $\vect{z}$.
We write
\bea
h_r(\vect{y},t)=A_0 t+B
              +\frac{M_r}{| \,\vect{y}-\vect{y}_r|^{n_r-2}},\quad
%and
h_s(\vect{z})=C
              +\frac{M_s}{|\,\vect{z}-\vect{z}_s|^{n_s-2}},
\eea
(see Eq. (\ref{2:solution:Eq})).
We need to assume either $M_r= 0$ or $M_s= 0$,
since in our construction the position of the brane world
is specified by a single coordinate $|\vect{y}|$ or $|\vect{z}|$.
For $M_r= 0$, we set $\xi= |\vect{z}|$ and $n_s=n$ with $\vect{z}_s=0$,
while for $M_s=0$ $\xi= |\vect{y}|$ and $n_r=n$ with $\vect{y}_r=0$.

\item{Case (III)}

Here $h_r$ and $h_s$ depend on $\,\vect{y}$ and $\vect{v}$.
We again write
\bea
h_r(\vect{y},t)=A_0 t+B
              +\frac{M_r}{|\,\vect{y}-\vect{y}_r|^{n_r-2}},\quad
h_s(\vect{v})=C
              +\frac{M_s}{|\,\vect{v}-\vect{v}_s|^{n_s-2}},
\eea
(see Eq. (\ref{3:solution:Eq})).
Similarly we need to set either $M_r= 0$ or $M_s=0$.
For $M_r= 0$, we set $\xi= |\vect{v}|$ and $n_s=n$ with $\vect{v}_s=0$,
while
for $M_s=0$, $\xi= |\vect{y}|$ and $n_r=n$ with $\,\vect{y}_r=0$.

\end{enumerate}

Let us illustrate a D3-D1 solution of the case (I), where the 
D3 brane is time dependent.
Under our assumptions, after compactifying the trivial $\Ysp_1$ directions,
the Einstein-frame metric is given by
\bea
ds^2&=&h_r^{-\frac{3}{7}}(\vect{z},t)h_s^{-\frac{6}{7}}(\vect{z})
\Big(
-dt^2+ h_s(\vect{z})\delta_{mn} dv^m dv^n
+h_r(\vect{z},t)h_s(\vect{z})\delta_{ab}dz^a dz^b
\Big),
\eea
where $v^m$ and $z^a$ are the coordinates of three- and five-dimensional
Euclidean spaces.
Now, we identify
$\xi=|\vect{z}|$, $\{\theta\}=
\text{angular part of} \,\{z\}$ and $\{{\cal X}\}=\{v\}$.
Then, functions in \eqref{metric_ge} read
$a=h_r^{-\frac{3}{14}}h_s^{\frac{1}{14}},
d=h_r^{-\frac{3}{14}}h_s^{-\frac{3}{7}},$ and 
$f=h_r^{\frac{2}{7}} h_s^{\frac{1}{14}}$,
respectively.
%g= \xi h_r^{\frac{2}{7}}$
%h_s^{\frac{1}{14}}$.

\subsection{Properties of brane world near the singularity}

\begin{figure}[h]
\begin{center}
\includegraphics[width=6cm,clip]{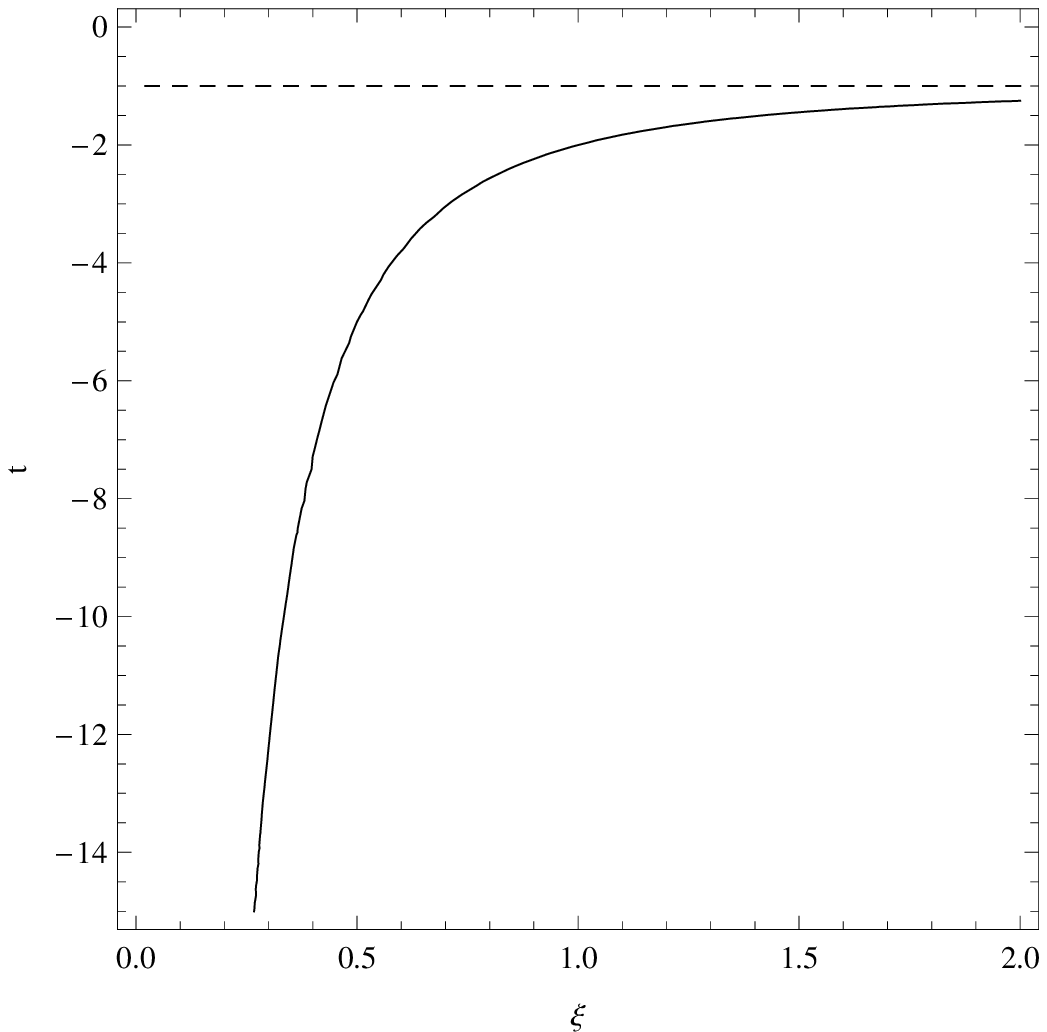}
\hskip 1cm
\includegraphics[width=6cm,clip]{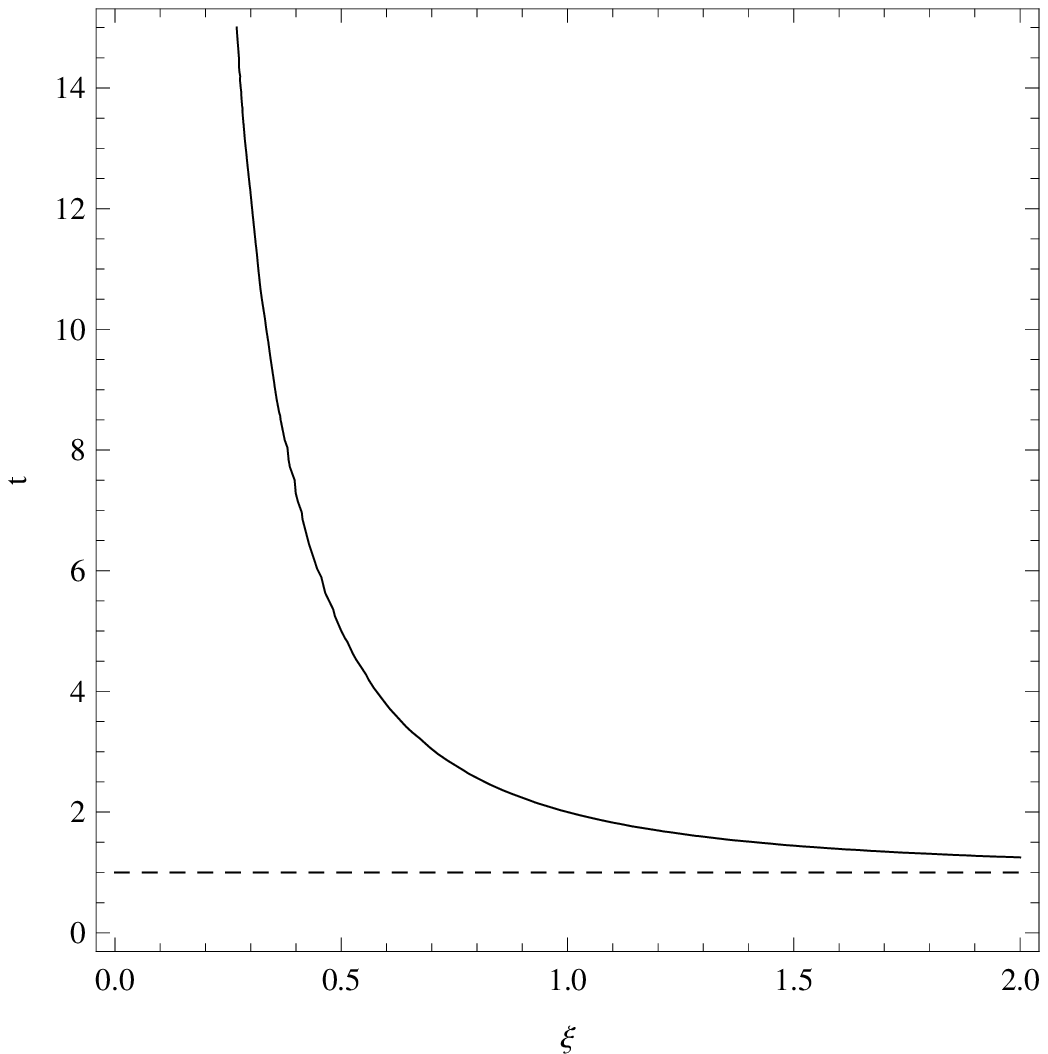}\\
(a) $A_0>0$ \hskip 4.5cm
(b) $A_0<0$
\caption{\baselineskip 12pt
Curves in the figures
show the level of $h_r=0$ (solid curves) in the case of $n=3$.
We set $|A_0|=1$.
The horizontal and vertical axes show
$\xi$ and $t$, respectively
The dashed lines denote the critical times.
(a) For $A_0>0$, the size of
the spacetime domain (the left side of the solid curve)
 expands and diverges at $t=-1$.
(b) For $A_0<0$, at $t=+1$ singularity appears at the infinity and then
the spacetime domain (the left side of the solid curve)
shrinks as the time progresses.
}
\label{cp}
\end{center}
\end{figure}

We discuss the properties of singularity at $h_r=0$,
which arises because of the time dependence.
We assume that $B>0$ and $M_r\geq 0$.
We also assume that $C>0$ and $M_s\geq 0$ in $h_s$, so that no singularity
other than the brane at $\xi=0$ appears from $h_s$.
In the case $n>2$, the dynamics of the spacetime is changed
at the critical time $t=-\frac{B}{A_0}$.
For $M_r>0$ and $A_0>0$, in the infinite past $t\to-\infty$
the regular spatial region is small.
The spatial region gradually expands
and spreads to the infinity at $t=-B/A_0$.
Subsequently, the spacetime is regular except at $\xi=0$.
For $A_0<0$, initially spacetime is regular except at the brane.
But at $t=\frac{B}{|A_0|}$, a singularity appears at the spatial infinity, and
then the spatial domain shrinks as the time evolves. See Fig. \ref{cp}.
In the case of $n=1$, for $M_r>0$ and $A_0>0$,
the spacetime is defined for $\xi>-\frac{1}{M_r}(A_0t+B)$,
while for $A_0<0$ it can be done for $\xi<\frac{1}{M_r}(|A_0|t-B)$.
For $M_r=0$, a spacelike singularity appears at $t=-\frac{B}{A_0}$.
For $A_0>0$, spacetime can be defined for $t>-\frac{B}{A_0}$
while for $A_0<0$ it can be done for $t<\frac{B}{|A_0|}$.

We embed the brane world and discuss the dynamics near the singularity.
In Tables \ref{table_26}-\ref{table_28}, we have classified the future
singularities of brane worlds for $A_0<0$.
The behavior of past singularities can be discussed for $A_0>0$.
In the tables, for example,
``D$m$-D$n$'' denotes the intersecting D-branes,
where the first D$m$-brane has the time dependence.
Also ``($A$,$B$)'' represents the directions of
the ordinary three-space and the bulk, respectively.
These rules are also applied to the subsequent tables.

%%%%%%%%%%%%%%%%%%%%%%%%%%%%%%%%%%%%%%%%%
%%%%%%%%%%%%%%%%%%%%%%%%%%%%%%%%%%%%%%%%%%%%%%

For $n>2$, in the metric (\ref{metric_ge})
$d^2\propto h_r^{q-1}$ ($f^2\propto h_r^{q}$ or $f^2\propto h_r^{q-1}$),
where $0<q<1$. % is a fraction.
From \eqref{normalization},
$\frac{dt}{d\tau} \sim  h_r^{\frac{1-q}{2}}\to 0$ as $h_r\to 0$,
which implies that for $A_0<0$ the brane world
cannot reach the future singularity within a finite proper time.
Note that for $f^2\propto h_r^q$ the brane world
trajectory is always timelike as $h_r\to 0$
while for $f^2\propto h_r^{q-1}$ it becomes
timelike or null in approaching singularity.
Conversely, for $A_0>0$, the universe is born in the infinite past
in terms of the proper time.
On the other hand,
the solutions listed in Table~\ref{table_29} give $q>1$,
and hence $\frac{dt}{d\tau}
\sim  h_r^{\frac{1-q}{2}}\to \infty$ as $h_r\to 0$.
Thus, the brane world can reach the singularity
within a finite proper time.
Some of the $n=1$ solutions are in this class.

\subsection{Cosmological equations}

Now we derive the cosmological equations in the brane world
through the junction conditions.
Under the $Z_2$ symmetry,
they are given by
\bea
\label{eq_junc}
K^{\hat A}{}_{\hat B}-\delta^{\hat A}{}_{\hat B}
K^{\hat C}{}_{\hat C}
=-\frac{1}{2}\kappa^2 S^{\hat A}{}_{\hat B},
\eea
where
$K_{\hat A\hat B}:=q_{\hat A}^{A}q_{\hat B}^B
\nabla_{A} n_B$
is the extrinsic curvature tensor,
and
$S^{\hat A}{}_{\hat B}$
is the energy-momentum tensor of the brane world matter.
$n_A$ and $q_{\hat A}^A$ are the unit normal vector
and projection tensor to the brane world, respectively,
The hatted indices $\{\hat A\}$ run over the brane world directions.
The nonvanishing $(i,j)$, $(a,b)$, and $(\tau,\tau)$-components of
the left-hand side of Eq.~(\ref{eq_junc}) are related to
the pressure in the ordinary 3-spatial direction,
that in the $(n-1)$-sphere
and the energy density on the brane world.

From now on, we focus on the (square of the) $(\tau,\tau)$ component,
which is given by
\begin{eqnarray}
\label{friedmann}
&&1+\frac{\dot{a}^2}{\alpha_\xi^2}
-\frac{\alpha_t^2
       \big(\alpha_\xi^4-\alpha_t^2 \dot{a}^2
       -\alpha_\xi^2(\alpha_t^2-3\dot{a}^2)
       \big)  }
      {\alpha_\xi^2\big(\alpha_\xi^2-\alpha_t^2\big)^2}
\pm
\frac{2\alpha_\xi\alpha_t \dot{a}
\sqrt{\alpha_\xi^2-\alpha_t^2+\dot{a}^2}}
     {(\alpha_\xi^2-\alpha_t^2)^2}
\nonumber \\
&=&\frac{1}{\big(3\frac{\alpha_\xi}{a}
+(n-1)\frac{\gamma_{\xi}}{f\xi}\big)^2}
\left\{
\frac{1}{2}\epsilon\kappa^2\rho
-\Big(3\frac{\alpha_t}{f}+(n-1)\frac{\gamma_t}{d}\Big)
\frac{\alpha_\xi \dot{a}\pm \alpha_t
\sqrt{\alpha_\xi^2-\alpha_t^2+\dot{a}^2}}
     {\alpha_\xi^2-\alpha_t^2}
\right\}^2,
\end{eqnarray}
where $\rho$ is the energy density in the brane world.
 %used $\dot{a}=a_{,t}\dot{t}+a_{,\xi}\dot{\xi}$ and
For convenience, we have introduced
\begin{eqnarray}
\alpha_\xi:=\frac{a_{,\xi}}{f},\quad
\alpha_t:=\frac{a_{,t}}{d},\quad
\gamma_\xi:=1+\frac{\xi f_{,\xi}}{f}\quad
\gamma_t:=\frac{\xi f_{,t}}{d}.
\end{eqnarray}
$\epsilon=+1$ ($-1$) denotes the normal vector pointing
the increasing (decreasing) $\xi$ direction.
When the spacetime is approximately static,
i.e., $|\alpha_\xi|\gg |\alpha_t|$ and $|\gamma_\xi|\gg |\gamma_t|$,
Eq. (\ref{friedmann}) reduces to
\begin{eqnarray}
\label{static_friedmann}
\frac{\dot{a}^2}{a^2}
\approx \frac{\kappa^4\rho^2}
      {4\big(3+(n-1)\frac{\gamma_\xi a}{\alpha_\xi f\xi}\big)^2}
-\frac{\alpha_\xi^2}{a^2},
\end{eqnarray}
where ``$\approx$'' becomes ``$=$'' when $\alpha_t= \gamma_t=0$.
When the time dependence dominates the system,
i.e.,
$|\alpha_\xi|\ll |\alpha_t|$ and $|\gamma_\xi|\ll |\gamma_t|$,
Eq. (\ref{friedmann}) reduces to
\begin{eqnarray}
\frac{\dot{a}^2}{a^2}
\approx \frac{\kappa^4\rho^2}
      {4\big(3+(n-1)
     \frac{\gamma_t a}{\alpha_t f\xi}\big)^2}
+\frac{\alpha_t^2}{a^2},
\label{dyn}
\end{eqnarray}
where ``$\approx$''  becomes ``$=$'' when $\alpha_{\xi}=0$
and $\gamma_{\xi}=1$.

Furthermore,
we integrate the energy density over the $(n-1)$-sphere in the brane world
$\rho_{{\rm 4D}}= \xi^{n-1} f^{n-1} V_{n-1} \rho$,
where $V_{n-1}:=\frac{2\pi^{n/2}}{\Gamma(n/2)}$
is the surface area of the unit $(n-1)$-sphere.
Then, we decompose the energy density
into the constant part $\sigma$ and
the time dependent one $\delta \rho$:
$\rho_{4D}:=\sigma+\delta \rho$.
%By ``low energy limit'' we mean $\delta\rho\ll \sigma$.
In both of the above limiting cases, at the low-energy scale 
($\delta\rho\ll \sigma$),
Eq.~(\ref{static_friedmann}) reduces to
\bea
\frac{\dot{a}^2}{a^2}
\approx
 \left\{ \begin{array}{ll}
 \frac{8\pi G_{\rm eff}}{3}\delta\rho
+\frac{1}{3}\Lambda^{(\xi)}_{\rm eff}(\tau)
+O(\delta \rho^2)
& (\text{static limit})
\\
%\frac{\dot{a}^2}{a^2}
%\approx
\frac{8\pi G_{\rm eff}}
      {3}
\delta \rho
+\frac{1}{3}\Lambda^{(t)}_{\rm eff}(\tau)
+O(\delta \rho^2)
& (\text{time-dependent limit})
\\
\end{array}
\right.
\label{td2}
\eea
where the effective gravitational constant is given by
\begin{eqnarray}
\label{qs3}
\frac{8\pi G_{\rm eff}}{3}
:=\left\{ \begin{array}{ll}
\frac{\kappa^4\sigma}
{2V_{n-1}^2 (\xi f)^{2n-2}\big(3+(n-1)
(\frac{1}{\xi}+\frac{f_{,\xi}}{f})\frac{a}{a_{,\xi}}
\big)^2}
& (\text{static limit}) \\
\frac{\kappa^4\sigma}
    {2V_{n-1}^2(\xi f)^{2n-2}\big(3
+(n-1)
\frac{f_{,t}a}{fa_{,t}}\big)^2}
& (\text{time-dependent limit})
\end{array} \right.
\end{eqnarray}
respectively.
$\Lambda^{(\xi)}_{\rm eff}(\tau)$ and $\Lambda^{(t)}_{\rm eff}(\tau)$
give the nonstandard contributions.
The necessary condition to obtain a realistic low-energy cosmology
is that the effective gravitational coupling in Eq. (\ref{qs3})
remains (almost) constant during the cosmological evolution.
%{\bf
It is clear that in the definition of $G_{\rm eff}$, Eq. (\ref{qs3}),
the motion of the $(n-1)$ compact dimensions on the brane world is taken in account.
Its effect makes it difficult to obtain a constant $G_{\rm eff}$.
Nevertheless,
as we will discuss later, we will find a few examples in which
the constant $G_{\rm eff}$ is obtained.
%}

%{\bf
Before closing this subsection,
we briefly mention the other requirement for the recovery
of the four-dimensional gravity,
i.e., the localizability of the graviton zero mode.
This issue is independent of the number of extra dimensions $n$.
Although to clarify it
we have to investigate the perturbations,
at least we need to require the
finiteness of the bulk volume off the brane world.
In our solutions, in general it is not ensured.
Then, it is necessary to add, for example,
the cut-off second brane world.
In this case, the gravity in the brane world may
not coincide with the ordinary general relativity,
due to the degree of freedom associated with the interbrane distance.
For now, we leave this issue for a future study
and focus on the minimal requirement
that $G_{\rm eff}$ must be constant.
%}

%\subsection{Brane world and singularity}

\subsection{Effective gravitational coupling}

Finally, we discuss the behavior of
the effective gravitational coupling~(\ref{qs3})
and list the realistic brane world models.

%%%%%%%%%%%%%%%%%%%%%%%%%%%%%%%%%%%%%%%%%%%%%%%%%%%%%%%%%%%%%%%%%
\subsubsection{$n=1$}
%%%%%%%%%%%%%%%%%%%%%%%%%%%%%%%%%%%%%%%%%%%%%%%%%%%%%%%%%%%%%%%%%

The requirement of a constant $G_{\rm eff}$
is trivially satisfied
for $n=1$, with $G_{\rm eff}=\frac{\kappa^4\sigma}{48\pi }$.
The solutions which give $n=1$ brane world models
are listed in Table~\ref{table_30}.
We classify the solutions with $h_r=A_0 t+ B +M_r |\xi|$ and
$h_r=A_0 t+ B$ in (a) and (b) in the table, respectively.
Also in the right column of Tables~\ref{table_1}-\ref{table_18},
the original ten- or eleven-dimensional solutions which give these
brane worlds are listed.

In Table~\ref{table_30_2} we further classify $n=1$ brane worlds
in terms of the behavior near the future singularity for $A_0<0$.
We find that no brane world model
whose scale factor diverges within a finite time,
i.e., a big-rip singularity, is obtained.
Solutions in the category of ($a=0$,``Finite'')
provide the brane worlds which collapse
within a finite time.
But ever expanding brane worlds
are also obtained from the intersecting D-branes of the cases (I) and (II).
For $A_0>0$, from the Table \ref{table_30_2},
the behavior of the brane near the past singularity can
be discussed.
In particular, the solutions of ($a=0$, Finite) in Table
\ref{table_30_2} provides the brane worlds with initial big bang
singularities in the finite past.

Before closing this part, it is important to summarize the property of
brane world models of $n=1$ in each case.
First, it is clear that in our treatment
M-branes do not provide
such a brane world model
in all the cases (I)-(III).
Concerning the ten-dimensional solutions,
we summarize their properties below:
\begin{description}

\item{\it Case (I):}

A remarkable property of this case is that
all the $n=1$ brane world models are obtained only from
the intersecting D-branes.
By definition, the bulk direction is always $\Zsp$.
In addition, all of models are of type (a) (see Table~\ref{table_30}).
In approaching the future $h_r=0$ singularity,
the model of D5-D7($\Ysp_2$,$\Zsp$) gives
an ever expanding universe, while
the model of D5-D7($\Ysp_1$,$\Zsp$) provides
an ever contracting universe.
All the rest give universes collapsing within the finite time.

\item{\it Case (II):}

All classes of solutions can provide
the $n=1$ brane world models.
This is in part because some of the case (II) solutions
correspond to particular cases of case (I).
The bulk direction can be either $\Ysp_1$ or Z.
For the models of type (a) $\Ysp_1$ is the bulk direction,
while for those of type (b) $\Zsp$ is the bulk direction.
In approaching the future $h_r=0$ singularity,
the model of D5-D7($\tilde \Xsp$,$\Zsp$) gives
an ever expanding universe, while
the models of
D5-D7($\Ysp_1$,$\Zsp$),
D3-D1($\Zsp$,$\Ysp_1$) and
NS5-D4($\Zsp$,$\Ysp_1$)
provide ever contracting universes.
All the rest give universes collapsing within the finite time.

\item{\it Case (III):}

The $n=1$ brane world models are obtained only from
the solutions including an NS5-brane.
The bulk direction can be either $\Ysp_1$ or $\Ysp_2$.
For the models of type (a) $\Ysp_1$ is the bulk direction,
while for those of type (b) $\Ysp_2$ is the bulk direction.
In approaching the future $h_r=0$ singularity,
the model of
NS5-D7($\Ysp_1$,$\Ysp_2$)
provides an ever contracting universe.
All the rest give universes collapsing within the finite time
and an ever expanding universe is not realized.

\end{description}

\subsubsection{$n>2$}
%%%%%%%%%%%%%%%%%%%%%%%%%%%%%%%%%%%%%%%%%%%%%%%%%%%%%%%%%%%%%%%%%

We consider the case of $n>2$.
First, for simplicity, we discuss the case of the completely static solutions
$\alpha_t=\gamma_t=0$.
In this case, there is no future or past singularity.
We do not find examples that the effective gravitational coupling
approaches a constant.
In the opposite limit $\xi\to \infty$, we find that
there are examples with constant $G_{\rm eff}$ in cases (I) and (III),
which are listed in Table~\ref{table_31}. From the table,
we see that for case (I) the four models of $n=3$, i.e.,
D5-D3 ($\Ysp_2$,$\Zsp$),
NS5-D3($\Ysp_2$,$\Zsp$),
D3-D5 ($\Ysp_1$,$\Zsp$),
D3-NS5($\Ysp_1$,$\Zsp$),
and
for the case (III)
D3-D7((${\tilde \Xsp}$,$\Ysp_2$),$\Ysp_1$) model
of $n=6$ can be realistic.
In case (I), we have to assume both $M_r\neq 0$ and $M_s\neq 0$.
In the right column of Tables \ref{table_1}-\ref{table_13},
the original ten-dimensional solutions which give these models
are indicated.
Note that the static M-brane solutions do not give realistic models.
%{\bf
Although in our model there are many solutions and ways of compactification,
the number of realistic solutions is rather small.
This is because the compact $(n-1)$ dimensions in the brane world
evolve with time and make $G_{\rm eff}$ in Eq. (\ref{qs3})
time dependent even in the static solutions.
%}

By allowing the time dependence
$\alpha_t\neq 0$ and $\gamma_t\neq 0$,
in the near-brane limit $\xi\to 0$, 
the static approximation is still valid,
namely
$|\alpha_t|\ll |\alpha_{\xi}|$ and
$|\gamma_{t}|\ll |\gamma_{\xi}|$.
But in the $\xi\to\infty$ limit,
the time dependence now dominates the system
since $|\alpha_t|\gg |\alpha_{\xi}|$ and
$|\gamma_{t}|\gg |\gamma_{\xi}|$.
Then, in both the limits of $\xi\to 0$ and $\xi\to \infty$,
there is no example of the constant gravitational coupling.
Therefore, for $n>2$, it is impossible to construct the realistic brane
world models from our solutions.

%%%%%%%%%%%%%%%%%%%%%%%%%%%%%%%%%%%%%%%%%%%%%%%%%%%%%%%%
%\subsubsection{ Purely time dependent solutions of $n>2$}

%The pure time dependent solutions are obtained by setting $M_r=M_s=0$
%for each class of the cases I-III.
%For any $n>1$
%there is no solution
%which provide a constant $G_{\rm eff}$.

%======================================%
%<<<<<<<<<<<<< SECTION 5 >>>>>>>>>>>>>>%
%======================================%

%T1>Discussions
\section{Discussions}
  \label{sec:discussions}

In this paper, we have derived intersecting dynamical brane solutions
and discussed their dynamics in the ten- and eleven-dimensional supergravity models.
These solutions are obtained by replacing a
constant $A$ in the warp factor $h=A+h_1(y)$ of
a supersymmetric solution by a function $h_0(x)$ of the
coordinates $x^\mu$
\cite{Gibbons:2005rt, Kodama:2005fz,Binetruy:2007tu, Maeda:2009zi}.
Our solutions can contain only one function depending on both time as well
as overall or relative transverse space coordinates.
In particular, the solutions in Sec.~\ref{sub:II} tell us that
the brane which depends on overall transverse coordinate can be
extended to the time dependent case.
It is possible to get the dynamical intersecting brane solutions
which obey the intersection rule $\chi=-2$ different from the usual one,
as we have discussed in Sec.~\ref{sub:III}.

%{\bf
We have used the intersection rules to find the cosmological
solution because it is not so easy to find it analytically
without their rules.
The intersection rules have led to the functions $h_r$ and $h_s$
which can be written by linear combinations of terms depending on both
coordinates of worldvolume and transverse space.
%}
This feature is expected to be shared by a wide class
of supersymmetric solutions beyond the examples considered
in the present paper, because the result has been obtained
by analyzing the general structure of solutions for warped
compactification with field strength of the ten- or eleven-dimensional
supergravities under ansatz that is natural to include
supersymmetric solutions as a special case.
We have showed that these solutions give a FLRW universe if we regard
the homogeneous and isotropic part of the brane worldvolumes
as our spacetime. Unfortunately, the power of the scale factor is
so small that the solutions of field equations cannot give a
realistic expansion law. This means that we have to consider additional
matter on the brane in order to get a realistic expanding universe.
As the number $p_r$ or $p_s$ increases, the power of the scale factor
becomes large. We find that the intersection with 
D6-brane in ten-dimensional
theory gives the fastest expansion of our universe because the
three-dimensional spatial space of our universe stays in the
transverse space to the D6-brane.
Though the power of the scale factor for the transverse space in
solutions with the D7- or D8-branes is larger than those with the D6-brane,
the number of the transverse space to these branes is less than three.
Hence, these solutions cannot provide the isotropic universe if
we assume that the transverse space to the brane is the part of
our universe.

%{\bf
The solutions we have obtained may give some moduli instabilities
because of the flat direction of the moduli potential in
the lower-dimensional effective theories after compactifications
\cite{Kodama:2005fz, Kodama:2005cz, Maeda:2009zi, Minamitsuji:2010fp}.
Such instability will grow unless the global or local
minimum of the potential can be produced by some correction in the
effective theory.
%}

The dynamical solutions contain only one function depending on both
time and transverse space coordinates. One possible reason for
this is that the ansatz concerning the structure of the
$D$-dimensional metric is too restrictive. However, a recent study of
similar systems shows that it is possible to obtain solutions
with each function depending on both time and transverse
space coordinates (see \cite{Gibbons:2009dr} for recent discussion).
It is interesting to examine if our solutions can be extended
to more general solutions by relaxing the assumptions of the field ansatz.

%%%%%%%%%%%%%% Comments on brane world models %%%%%%%%%%%%%%%%%%%%%%%
Finally, we have constructed the brane world models from our solutions.
This approach makes it clear how the ordinary four-dimensional matter
contributes to the cosmology.
In our approach,
we first compactify the trivial spatial directions in
a given ten- or eleven-dimensional spacetime
and then move to the Einstein frame.
This approach gives a way of regularization of the brane source
to put matter there.
For a brane with higher codimensions
we have applied the cut-copy-paste method. % (see Fig. \ref{BW}).
We then need to integrate over the angular dimensions in the brane worldvolume
to define the effective four-dimensional quantities.
For a codimension-one brane, we need just the copy and paste.
For our prescription to work, we have restricted our solutions.
In particular, we have chosen the parameters of
ten- or eleven-dimensional solutions that after compactifying
the trivial spatial dimensions the extra dimensions
become spherically symmetric with respect to a single brane.
We have also classified the singularity in the time dependent solutions,
and discussed the behaviors of the brane world universe around it.

Then, we have derived the effective gravitational equations
via the junction condition.
The necessary condition to obtain a realistic cosmological model
is that the effective gravitational constant must approach constant.
A brane world model obtained from a codimension-one brane
can automatically realize a constant gravitational coupling.
The intersecting M-branes could not provide such models
for all types of brane intersection.
The existence of some ten-dimensional solutions which provide
such brane world models
crucially depend on the types of the brane intersection.
In terms of the behavior around the singularity,
an ever expanding universe cannot be obtained
from the solutions where the metric does not depend on
the overall transverse space.
Concerning the models constructed from higher-codimensional branes,
for the purely static case, we have found a few solutions
where the effective gravitational coupling approaches constant
in the far brane limit.
In contrast, however, for the generic time dependent solutions,
we did not find such examples.

%\begin{thebibliography}{99}

%\begin{table}[ht]
\begin{table}[p]
\caption{\baselineskip 14pt
Intersections of two M-branes in the metric
\eqref{b:metric-a:Eq} and \eqref{b:metric-mr:Eq}.
%Here is no difference whichever of the two M2's or M5's is time dependent.
Whichever of the two M2's or M5's is time dependent does not 
make any difference. 
In the right column of the Tables~\ref{table_1}-\ref{table_13},
 ``BW'' denotes the solutions which provide realistic brane world models.
From the solutions in the list, no realistic brane world model is obtained.}
\label{twoM}
{\scriptsize
\begin{center}
\begin{tabular}{|c||c|c|c|c|c|c|c|c|c|c|c|c||c||c|c|c|c|}
\hline
Case&&0&1&2&3&4&5&6&7&8&9&10& & $\tilde{\Msp}$ & $\lambda(\tilde{\Msp})$
& $\lambda_{\rm E}(\tilde{\Msp})$& BW
\\
\hline
(I), (II) &M2 & $\circ$ &&& $\circ$ & $\circ$ &&&&&&&$\surd$
& $\Ysp_1$ \& $\Zsp$ & $\lambda(\Ysp_1)$=1/4 & $\lambda_{\rm E}(\Ysp_1)=
\frac{-3+d_3}{-12+d_2+2d_3+d_4}$
&%$\circ$
\\
\cline{3-13}
M2-M2&M2 & $\circ$ & $\circ$ & $\circ$ &&&&&& & &&
&& $\lambda(\Zsp)$=1/4 & $\lambda_{\rm E}(\Zsp)=
\frac{-3+d_3}{-12+d_2+2d_3+d_4}$
&%$\circ$
\\
\cline{3-13}
&$x^N$ & $t$ & $y^1$ & $y^2$ & $v^1$ & $v^2$ & $z^1$ & $z^2$ & $z^3$
& $z^4$ & $z^5$ & $z^6$ &
&& &
&%$\circ$
\\
\hline
\hline
(I), (II) &M2 & $\circ$ & $\circ$ &&  &   &  & $\circ$ &&&&& $\surd$
& $\Ysp_1$ \& $\Zsp$ & $\lambda(\Ysp_1)$=1/4 &
$\lambda_{\rm E}(\Ysp_1)=\frac{-3+d_1+d_3}{-12+2d_1+d_2+2d_3+d_4}$
&%$\circ$
\\
\cline{3-13}
M2-M5&M5 & $\circ$ & $\circ$ & $\circ$ &$\circ$& $\circ$ &$\circ$&& &
& &&& &  $\lambda(\Zsp)$=1/4 &
$\lambda_{\rm E}(\Zsp)=\frac{-3+d_1+d_3}{-12+2d_1+d_2+2d_3+d_4}$
&%$\circ$
\\
\cline{3-13}
&$x^N$ & $t$ & $x$ & $y^1$ & $y^2$ & $y^3$ & $y^4$ & $v$ & $z^1$
& $z^2$ & $z^3$ & $z^4$ &
&& &
&%$\circ$
\\
%\cline{2-17}
\hline
(I), (II) &M2 & $\circ$ & $\circ$ & $\circ$ &   &   &   &&&&&&
&  $\tilde{\rm X}$ \& $\Ysp_2$& $\lambda(\Ysp_2)=-1/5$ &
$\lambda_{\rm E}(\Ysp_2)=\frac{3-d_2-d_4}{-15+2d_1+d_2+2d_3+d_4}$
&%$\circ$
\\
\cline{3-13}
M2-M5&M5 & $\circ$ &$\circ$ & & $\circ$ &$\circ$ & $\circ$& $\circ$ & & & &&
 $\surd$ & or & $\lambda(\Ysp_1)$=2/5 &
$\lambda_{\rm E}(\Ysp_1)=\frac{-6+d_1+d_3}{-15+2d_1+d_2+2d_3+d_4}$
&%$\circ$
\\
\cline{3-13}
&$x^N$ & $t$ & $x$ & $y$ & $v^1$ & $v^2$ & $v^3$ & $v^4$ & $z^1$
& $z^2$ & $z^3$ & $z^4$ &
&$\Ysp_1$ \& $\Zsp$ & $\lambda(\Zsp)$=2/5 &
$\lambda_{\rm E}(\Zsp)=\frac{-6+d_1+d_3}{-15+2d_1+d_2+2d_3+d_4}$
&%$\circ$
\\
\hline
\hline
(I), (II) &M5 & $\circ$ & $\circ$ & $\circ$ & $\circ$ &&
& $\circ$ & $\circ$ &&&&$\surd$
& $\tilde{\rm X}$ \& $\Ysp_2$ & $\lambda(\tilde{\rm X})=-1/5$ &
$\lambda_{\rm E}(\tilde{\rm X})=
\frac{3-d_2-d_4}{-15+2d_1+d_2+2d_3+d_4}$
&%$\circ$
 \\
\cline{3-13}
M5-M5&M5 & $\circ$ & $\circ$ & $\circ$ & $\circ$ & $\circ$ & $\circ$ &&& & &&
& or  &$\lambda(\Ysp_2)=-1/5$ &
$\lambda_{\rm E}(\Ysp_2)=
\frac{3-d_2-d_4}{-15+2d_1+d_2+2d_3+d_4}$
&%$\circ$
\\
\cline{3-13}
&$x^N$ & $t$ & $x^1$ & $x^2$ & $x^3$ & $y^1$ & $y^2$ & $v^1$
& $v^2$ & $z^1$ & $z^2$ & $z^3$ &
& $\Ysp_1$ \& $\Zsp$  & $\lambda(\Ysp_1)=\lambda(\Zsp)=\frac{2}{5}$ &
$\lambda_{\rm E}(\Ysp_1)=\lambda_{\rm E}(\Zsp)=\frac{-6+d_1+d_3}
{-15+2d_1+d_2+2d_3+d_4}$
&%$\circ$
\\
%\cline{2-17}
\hline
(III)&M5 & $\circ$ & $\circ$ &&&&&$\circ$ & $\circ$ & $\circ$ & $\circ$ &&
$\surd$& $\tilde{\Xsp}$ \& $\Ysp_2$ & $\lambda(\tilde{\Xsp})=-1/5$ &
 $\lambda_{\rm E}(\tilde{\Xsp})=
\frac{3-d_2-d_4}{-15+2d_1+d_2+2d_3+d_4}$
&%$\circ$
\\
\cline{3-13}
M5-M5&M5 & $\circ$ & $\circ$ & $\circ$ & $\circ$ & $\circ$ & $\circ$
&&&&&&
& or  & $\lambda(\Ysp_2)=-1/5$ & $\lambda_{\rm E}(\Ysp_2)=
\frac{3-d_2-d_4}{-15+2d_1+d_2+2d_3+d_4}$
&%$\circ$
\\
\cline{3-13}
&$x^N$ & $t$ & $x$ & $y^1$ & $y^2$ & $y^3$ & $y^4$ & $v^1$ & $v^2$ & $v^3$
& $v^4$ & $z$ &
& $\Ysp_1$ \& $\Zsp$ & $\lambda(\Ysp_1)=\lambda(\Zsp)=\frac{2}{5}$
& $\lambda_{\rm E}(\Ysp_1)=\lambda_{\rm E}(\Zsp)=\frac{-6+d_1+d_3}
{-15+2d_1+d_2+2d_3+d_4}$
&%$\circ$
\\
\hline
\end{tabular}
\end{center}
}
\label{table_1}
\end{table}

%\begin{table}[ht]
\begin{table}[p]
\caption{\baselineskip 14pt
Intersections of two D-branes with $p=0$ intersection in cases (I) and (II).
%Here is no difference whichever of the two D2's is time dependent.
Whichever of the two D2' is time dependent does not 
make any difference. 
In the right column, the notation $(A,B)$ corresponds to
the directions which give the ordinary 3-space ${\cal X}^i$
and the bulk $(\xi,\theta^a)$ of the brane world, respectively.
The D3-D1 solution can give a brane world model.
}
\label{twoDI-1}
{\scriptsize
\begin{center}
\begin{tabular}{|c||c|c|c|c|c|c|c|c|c|c|c||c||c|c|c|c|}
\hline
Branes&&0&1&2&3&4&5&6&7&8&9& & $\tilde{\Msp}$ & $\lambda(\tilde{\Msp})$
& $\lambda_{\rm E}(\tilde{\Msp})$
& BW
\\
\hline
&D0 &  $\circ$ &&&&&&&&&&$\surd$
& $\Ysp_1$ \& $\Zsp$ & $\lambda(\Ysp_1)$=1/9 & $\lambda_{\rm E}(\Ysp_1)=
\frac{1}{9-d_2-d_4}$
&%$\circ$
\\
\cline{3-12}
D0-D4&D4 & $\circ$ & $\circ$ & $\circ$  & $\circ$ & $\circ$
& & &&&&
&&$\lambda(\Zsp)$=1/9  & $\lambda_{\rm E}(\Zsp)=
\frac{1}{9-d_2-d_4}$
&%$\circ$
\\
\cline{3-12}
&$x^N$ & $t$ & $y^1$ & $y^2$ & $y^3$ & $y^4$ & $z^1$ & $z^2$
& $z^3$ & $z^4$ & $z^5$ &
&& &
&%$\circ$
\\
%\cline{1-17}
\hline
&D0 & $\circ$ &&&&&&&&&&
& $\Ysp_2$ or $\Zsp$ & $\lambda(\Ysp_2)=-3/13$ & $\lambda_{\rm E}(\Ysp_2)=
\frac{-3+d_4}{13-2d_3-d_4}$
&%$\circ$
\\
\cline{3-12}
D0-D4& D4 & $\circ$ & $\circ$ & $\circ$ & $\circ$ & $\circ$ &
& &&&&
$\surd$&& $\lambda(\Zsp)=5/13$ &$\lambda_{\rm E}(\Zsp)=
\frac{5-d_3}{13-2d_3-d_4}$
&%$\circ$
\\
\cline{3-12}
&$x^N$ & $t$ & $v^1$ & $v^2$ & $v^3$ & $v^4$ & $z^1$ & $z^2$
& $z^3$ & $z^4$ & $z^5$ &
&& &
&%$\circ$
\\
%\cline{1-17}
\hline
\hline
&D1 & $\circ$ &&&   &  $\circ$  &&&&&& $\surd$
& $\Ysp_1$ \& $\Zsp$ & $\lambda(\Ysp_1)$=1/5&
$\lambda_{\rm E}(\Ysp_1)=\frac{2-d_3}{10-d_2-2d_3-d_4}$
&%$\circ$
\\
\hline
\end{tabular}
\end{center}
}
\label{table_2}
\end{table}

%\begin{table}[ht]
\begin{table}[p]
\caption{\baselineskip 14pt
Intersections of two D-branes with $p=1$ intersection
in cases (I) and (II).
%Here is no difference whichever of the two D3's is time dependent.
Whichever of the two D3' is time dependent does not 
make any difference. 
No solution provides a realistic brane world model.
}
\label{twoDI-2}
{\scriptsize
\begin{center}
\begin{tabular}{|c||c|c|c|c|c|c|c|c|c|c|c||c||c|c|c|c|}
\hline
Branes&&0&1&2&3&4&5&6&7&8&9& & $\tilde{\Msp}$ & $\lambda(\tilde{\Msp})$
& $\lambda_{\rm E}(\tilde{\Msp})$& BW
\\
\hline
&D1 & $\circ$ & $\circ$ &&   &  &&&&&& $\surd$
& $\Ysp_1$ \& $\Zsp$ & $\lambda(\Ysp_1)$=1/5&
$\lambda_{\rm E}(\Ysp_1)=\frac{2-d_1}{10-2d_1-d_2-d_4}$
&%$\circ$
\\
\cline{3-12}
D1-D5&D5 & $\circ$ & $\circ$ & $\circ$ & $\circ$ &$\circ$& $\circ$
& & && &
& &  $\lambda(\Zsp)$=1/5&
$\lambda_{\rm E}(\Zsp)=\frac{2-d_1}{10-2d_1-d_2-d_4}$
&%$\circ$
\\
\cline{3-12}
&$x^N$ & $t$ & $x$ & $y^1$ & $y^2$ & $y^3$ & $y^4$ & $z^1$
& $z^2$ & $z^3$ & $z^4$ &
&& &
&%$\circ$
\\
%\cline{2-17}
\hline
&D1 & $\circ$ & $\circ$ & &   &   &&&&&&
& $\tilde{\Xsp}$ \& $\Ysp_2$ & $\lambda(\tilde{\Xsp})=-1/7$ &
$\lambda_{\rm E}(\tilde{\Xsp})=\frac{-2+d_4}{14-2d_1-2d_3-d_4}$
&%$\circ$
\\
\cline{3-12}
D1-D5&D5 & $\circ$ & $\circ$ & $\circ$ & $\circ$ & $\circ$& $\circ$
 & & & & &
 $\surd$ & or $\Zsp$ &  $\lambda(\Ysp_2)=-1/7$ &
$\lambda_{\rm E}(\Ysp_2)=\frac{-2+d_4}{14-2d_1-2d_3-d_4}$
&%$\circ$
\\
\cline{3-12}
&$x^N$ & $t$ & $x$ & $v^1$ & $v^2$ & $v^3$ & $v^4$ & $z^1$
& $z^2$ & $z^3$ & $z^4$ &
&& $\lambda(\Zsp)=3/7$ &
$\lambda_{\rm E}(\Zsp)=\frac{6-d_1-d_3}{14-2d_1-2d_3-d_4}$
&%$\circ$
\\
\hline
\hline
&D2 & $\circ$ & $\circ$ &&   &  & $\circ$ &&&&& $\surd$
& $\Ysp_1$ \& $\Zsp$ & $\lambda(\Ysp_1)$=3/11 &
$\lambda_{\rm E}(\Ysp_1)=\frac{3-d_1-d_3}{11-2d_1-d_2-2d_3-d_4}$
&%$\circ$
\\
\hline
\end{tabular}
\end{center}
}
\label{table_3}
\end{table}

%\begin{table}[ht]
\begin{table}[p]
\caption{\baselineskip 14pt
Intersections of two D-branes with $p=2$ intersection
in cases (I) and (II).
%Here is no difference whichever of the two D4's is time dependent.
Whichever of the two D4' is time dependent does not 
make any difference. 
The static D3-D5 solution of case (I) provides realistic brane world models.
}
\label{twoDI-3}
{\scriptsize
\begin{center}
\begin{tabular}{|c||c|c|c|c|c|c|c|c|c|c|c||c||c|c|c|c|}
\hline
Branes&&0&1&2&3&4&5&6&7&8&9& & $\tilde{\Msp}$ & $\lambda(\tilde{\Msp})$
& $\lambda_{\rm E}(\tilde{\Msp})$
& BW
\\
\hline
&D2 & $\circ$ & $\circ$ & $\circ$ &   &  &&&&&& $\surd$
&  $\Ysp_1$ \& $\Zsp$  & $\lambda(\Ysp_1)$=3/11 &
$\lambda_{\rm E}(\Ysp_1)=\frac{3-d_1}{11-2d_1-d_2-d_4}$
&%$\circ$
\\
\cline{3-12}
D2-D6&D6 & $\circ$ & $\circ$ & $\circ$ & $\circ$ &$\circ$& $\circ$
& $\circ$ & &&& & &  $\lambda(\Zsp)$=3/11 &
$\lambda_{\rm E}(\Zsp)=\frac{3-d_1}{11-2d_1-d_2-d_4}$
&%$\circ$
\\
\cline{3-12}
&$x^N$ & $t$ & $x^1$ & $x^2$ & $y^1$ & $y^2$ & $y^3$ & $y^4$
& $z^1$ & $z^2$ & $z^3$ &
&& &
&%$\circ$
\\
%\cline{2-17}
\hline
&D2 & $\circ$ & $\circ$ & $\circ$ &   &   &&&&&&
& $\tilde{\Xsp}$ \& $\Ysp_2$ & $\lambda(\tilde{\Xsp})=-1/15$ &
$\lambda_{\rm E}(\tilde{\Xsp})=\frac{d_4-1}{15-2d_1-2d_3-d_4}$
&%$\circ$
\\
\cline{3-12}
D2-D6&D6 & $\circ$ & $\circ$ & $\circ$ & $\circ$ & $\circ$& $\circ$
& $\circ$ & &&&
 $\surd$ & or $\Zsp$ &  $\lambda(\Ysp_2)=-1/15$ &
 $\lambda_{\rm E}(\Ysp_2)=\frac{d_4-1}{15-2d_1-2d_3-d_4}$
&%$\circ$
\\
\cline{3-12}
&$x^N$ & $t$ & $x^1$ & $x^2$ & $v^1$ & $v^2$ & $v^3$ & $v^4$
& $z^1$ & $z^2$ & $z^3$ &
&&$ \lambda(\Zsp)=7/15$ &
$\lambda_{\rm E}(\Zsp)=\frac{7-d_1-d_3}{15-2d_1-2d_3-d_4}$
&%$\circ$
\\
\hline
\hline
&D3 & $\circ$ & $\circ$ & $\circ$ &&  && $\circ$ &&&& $\surd$
& $\tilde{\Xsp}$ \& $\Ysp_2$ & $\lambda(\tilde{\Xsp})=-1/3$&
$\lambda_{\rm E}(\tilde{\Xsp})=\frac{d_2+d_4-4}{12-2d_1-d_2-2d_3-d_4}$
&$(\Ysp_1, \Zsp)$, I(s)
%$\circ$
\\
\hline
\end{tabular}
\end{center}
}
\label{table_4}
\end{table}

%\begin{table}[ht]
\begin{table}[p]
\caption{\baselineskip 14pt
Intersections of two D-branes with $p=3$ intersection
in cases (I) and (II).
%Here is no difference whichever of the two D5's is time dependent.
Whichever of the two D5' is time dependent does not 
make any difference. 
No solution provides a realistic brane world model.
}
\label{twoDI-4}
{\scriptsize
\begin{center}
\begin{tabular}{|c||c|c|c|c|c|c|c|c|c|c|c||c||c|c|c|c|}
\hline
Branes&&0&1&2&3&4&5&6&7&8&9& & $\tilde{\Msp}$ & $\lambda(\tilde{\Msp})$
& $\lambda_{\rm E}(\tilde{\Msp})$&
BW
\\
\hline
&D3 & $\circ$ & $\circ$ & $\circ$ & $\circ$ &  &&&&&& $\surd$
& $\tilde{\Xsp}$  & $\lambda(\tilde{\Xsp})=-1/3$ &
$\lambda_{\rm E}(\tilde{\Xsp})=\frac{d_2+d_4-4}{12-2d_1-d_2-d_4}$
&
%$\circ$
\\
\cline{3-12}
D3-D7&D7 & $\circ$ & $\circ$ & $\circ$ & $\circ$ & $\circ$&
$\circ$ & $\circ$ &  $\circ$ &&& & or
& $\lambda(\Ysp_1)=1/3$ &
$\lambda_{\rm E}(\Ysp_1)=\frac{4-d_1}{12-2d_1-d_2-d_4}$
&%$\circ$
\\
\cline{3-12}
&$x^N$ & $t$ & $x^1$ & $x^1$ & $x^3$ & $y^1$ & $y^2$ & $y^3$
& $y^4$ & $z^1$ & $z^2$ &
&$\Ysp_1$ \& $\Zsp$ & $\lambda(\Zsp)$=1/3 &
$\lambda_{\rm E}(\Zsp)=\frac{4-d_1}{12-2d_1-d_2-d_4}$
&%$\circ$
\\
%\cline{2-17}
\hline
&D3 & $\circ$ & $\circ$ & $\circ$ & $\circ$ &   &&&&&&
&$\tilde{\Xsp}$ \& $\Ysp_2$& $\lambda(\tilde{\Xsp})=0$ &
$\lambda_{\rm E}(\tilde{\Xsp})=\frac{d_4}{16-2d_1-2d_3-d_4}$
&%$\circ$
\\
\cline{3-12}
D3-D7&D7 & $\circ$ & $\circ$ & $\circ$ & $\circ$ & $\circ$& $\circ$
& $\circ$ &  $\circ$ &&&
 $\surd$ &  & $\lambda(\Ysp_2)=0$  &
 $\lambda_{\rm E}(\Ysp_2)=\frac{d_4}{16-2d_1-2d_3-d_4}$
&%$\circ$
\\
\cline{3-12}
&$x^N$ & $t$ & $x^1$ & $x^2$ & $x^3$ & $v^1$ & $v^2$ & $v^3$
& $v^4$ & $z^1$ & $z^2$ &
&& &
&%$\circ$
\\
\hline
\hline
&D4 & $\circ$ & $\circ$ & $\circ$ & $\circ$ &&&& $\circ$ &&& $\surd$
& $\tilde{\Xsp}$ \& $\Ysp_2$ & $\lambda(\tilde{\Xsp})=\lambda(\Ysp_2)=
\frac{-3}{13}$ &
$\lambda_{\rm E}(\tilde{\Xsp})=\lambda_{\rm E}(\Ysp_2)=
\frac{d_2+d_4-3}{13-2d_1-d_2-2d_3-d_4}$
&%$\circ$
\\
\hline
\end{tabular}
\end{center}
}
\label{table_5}
\end{table}

%\begin{table}[ht]
\begin{table}[p]
\caption{\baselineskip 14pt
Intersections of two D-branes with $p=4$ intersection
in cases (I) and (II).
%Here is no difference whichever of the two D6's is time dependent.
Whichever of the two D6' is time dependent does not 
make any difference. 
D5-D7 solutions of the cases (I) and (II) provide realistic brane world model.
}
\label{twoDI-5}
{\scriptsize
\begin{center}
\begin{tabular}{|c||c|c|c|c|c|c|c|c|c|c|c||c||c|c|c|c|}
\hline
Branes&&0&1&2&3&4&5&6&7&8&9& & $\tilde{\Msp}$ & $\lambda(\tilde{\Msp})$
& $\lambda_{\rm E}(\tilde{\Msp})$&
BW
\\
\hline
&D4 & $\circ$ & $\circ$ & $\circ$ & $\circ$ & $\circ$&&&&&& $\surd$
& $\tilde{\Xsp}$ & $\lambda(\tilde{\Xsp})=-3/13$ &
$\lambda_{\rm E}(\tilde{\Xsp})=\frac{d_2+d_4-3}{13-2d_1-d_2-d_4}$
&%$\circ$
\\
\cline{3-12}
D4-D8&D8 & $\circ$ & $\circ$ & $\circ$ & $\circ$ &$\circ$& $\circ$ & $\circ$ &
 $\circ$ & $\circ$ &&& $\Ysp_1$ \& $\Zsp$ & $\lambda(\Ysp_1)=5/13$&
$\lambda_{\rm E}(\Ysp_1)=\frac{5-d_1}{13-2d_1-d_2-d_4}$
&%$\circ$
\\
\cline{3-12}
&$x^N$ & $t$ & $x^1$ & $x^2$ & $x^3$ & $x^4$ & $y^1$ & $y^2$
& $y^3$ & $y^4$ & $z$ &
&& $\lambda(\Zsp)=5/13$ &
$\lambda_{\rm E}(\Zsp)=\frac{5-d_1}{13-2d_1-d_2-d_4}$
&%$\circ$
\\
%\cline{2-17}
\hline
&D4 & $\circ$ & $\circ$ & $\circ$ & $\circ$ & $\circ$ &&&&&&
& $\tilde{\Xsp}$ \& $\Ysp_2$ & $\lambda(\tilde{\Xsp})=1/17$ &
$\lambda_{\rm E}(\tilde{\Xsp})=\frac{d_4+1}{17-2d_1-2d_3-d_4}$
&%$\circ$
\\
\cline{3-12}
D4-D8&D8 & $\circ$ & $\circ$ & $\circ$ & $\circ$ & $\circ$&
$\circ$ & $\circ$ &  $\circ$ & $\circ$ &&
 $\surd$ & & $\lambda(\Ysp_2)=1/17$&
$\lambda_{\rm E}(\Ysp_2)=\frac{d_4+1}{17-2d_1-2d_3-d_4}$
&%$\circ$
\\
\cline{3-12}
&$x^N$ & $t$ & $x^1$ & $x^2$ & $x^3$ & $x^4$ & $v^1$ & $v^2$
& $v^3$ & $v^4$ & $z$ &
&& &
&%$\circ$
\\
\hline
\hline
&D5 & $\circ$ & $\circ$ & $\circ$ & $\circ$ & $\circ$ &&&& $\circ$ && $\surd$
& $\tilde{\Xsp}$ \& $\Ysp_2$ & $\lambda(\tilde{\Xsp})=\lambda(\Ysp_2)=
\frac{-1}{7}$ &
$\lambda_{\rm E}(\tilde{\Xsp})=\lambda_{\rm E}(\Ysp_2)=
\frac{d_2+d_4-2}{14-2d_1-d_2-2d_3-d_4}$
&$(\Ysp_1, \Zsp)$, I, II
%$\circ$
\\
\hline
\end{tabular}
\end{center}
}
\label{table_6}
\end{table}

%\begin{table}[ht]
\begin{table}[p]
\caption{\baselineskip 14pt
Intersections of two D-branes with $p=5$ intersection
in cases (I) and (II).
%Here is no difference whichever of the two D7's is time dependent.
Whichever of the two D7' is time dependent does not 
make any difference. 
No solution provides a realistic brane world model.
}
\label{twoDI-6}
{\scriptsize
\begin{center}
\begin{tabular}{|c||c|c|c|c|c|c|c|c|c|c|c||c||c|c|c|c|}
\hline
Branes&&0&1&2&3&4&5&6&7&8&9& & $\tilde{\Msp}$ & $\lambda(\tilde{\Msp})$
& $\lambda_{\rm E}(\tilde{\Msp})$&
BW
\\
\hline
&D6 & $\circ$ & $\circ$ & $\circ$ & $\circ$ & $\circ$& $\circ$ &
&&& $\circ$ & $\surd$
& $\tilde{\Xsp}$ \& $\Ysp_2$ & $\lambda(\tilde{\Xsp})=-1/15$ &
$\lambda_{\rm E}(\tilde{\Xsp})=\frac{d_2-1}{15-2d_1-d_2-2d_3}$
&%$\circ$
\\
\cline{3-12}
D6-D8&D8 & $\circ$ & $\circ$ & $\circ$ & $\circ$ &$\circ$& $\circ$ & $\circ$ &
 $\circ$ & $\circ$ &&& or $\Ysp_1$ &
 $\lambda(\Ysp_2)=-1/15$ &
$\lambda_{\rm E}(\Ysp_2)=\frac{d_2-1}{15-2d_1-d_2-2d_3}$
&%$\circ$
\\
\cline{3-12}
&$x^N$ & $t$ & $x^1$ & $x^2$ & $x^3$ & $x^4$ & $x^5$ & $y^1$
& $y^2$ & $y^3$ & $v$ &
&& $\lambda(\Ysp_1)=7/15$ &
$\lambda_{\rm E}(\Ysp_1)=\frac{7-d_1-d_3}{15-2d_1-d_2-2d_3}$
&%$\circ$
\\
%\cline{2-17}
\hline
&D6 & $\circ$ & $\circ$ & $\circ$ & $\circ$ & $\circ$ & $\circ$ & $\circ$
&&&&
& $\tilde{\Xsp}$ \& $\Ysp_2$ & $\lambda(\tilde{\Xsp})=1/17$ &
$\lambda_{\rm E}(\tilde{\Xsp})=\frac{d_2+1}{17-2d_1-d_2-2d_3}$
&%$\circ$
\\
\cline{3-12}
D6-D8&D8 & $\circ$ & $\circ$ & $\circ$ & $\circ$ & $\circ$&
$\circ$ &&  $\circ$ & $\circ$ & $\circ$ &
 $\surd$ & & $\lambda(\Ysp_2)=1/17$ &
$\lambda_{\rm E}(\Ysp_2)=\frac{d_2+1}{17-2d_1-d_2-2d_3}$
&%$\circ$
\\
\cline{3-12}
&$x^N$ & $t$ & $x^1$ & $x^2$ & $x^3$ & $x^4$ & $x^5$ & $y$
& $v^1$ & $v^2$ & $v^3$ &
&& &
&%$\circ$
\\
\hline
\hline
&D7 & $\circ$ & $\circ$ & $\circ$ & $\circ$ & $\circ$ & $\circ$ &
&&$\circ$ &$\circ$& $\surd$
& $\tilde{\Xsp}$ \& $\Ysp_2$ & $\lambda(\tilde{\Xsp})=0$ &
$\lambda_{\rm E}(\tilde{\Xsp})=\frac{d_2}{16-2d_1-d_2-2d_3}$
&%$\circ$
\\
\hline
\end{tabular}
\end{center}
}
\label{table_7}
\end{table}

%\begin{table}[ht]
\begin{table}[p]
\caption{\baselineskip 14pt
Intersections of two D-branes with $p=0$ intersection in case (III).
%Here is no difference whichever of
%the two D4's is time dependent.
Whichever of the two D4' is time dependent does not 
make any difference. 
No solution provides a realistic brane world model.
}
\label{twoDIII}
{\scriptsize
\begin{center}
\begin{tabular}{|c||c|c|c|c|c|c|c|c|c|c|c||c||c|c|c|c|}
\hline
Branes&&0&1&2&3&4&5&6&7&8&9& & $\tilde{\Msp}$ & $\lambda(\tilde{\Msp})$
& $\lambda_{\rm E}(\tilde{\Msp})$&
BW
\\
\hline
&D0 &  $\circ$ &&&&&&&&&&$\surd$
& $\Ysp_1$ \& $\Zsp$ & $\lambda(\Ysp_1)$=1/9 & $\lambda_{\rm E}(\Ysp_1)=
\frac{1}{9-d_2-d_4}$
&%$\circ$
\\
\cline{3-12}
D0-D8&D8 & $\circ$ & $\circ$ & $\circ$  & $\circ$ & $\circ$
& $\circ$ & $\circ$ & $\circ$ & $\circ$ &&
&& $\lambda(\Zsp)$=1/9 &
$\lambda_{\rm E}(\Zsp)=
\frac{1}{9-d_2-d_4}$
&%$\circ$
\\
\cline{3-12}
&$x^N$ & $t$ & $y^1$ & $y^2$ & $y^3$ & $y^4$ & $y^5$ & $y^6$
& $y^7$ & $y^8$ & $z$ &
&& &
&%$\circ$
\\
%\cline{1-17}
\hline
&D0 & $\circ$ &&&&&&&&&&
& $\Ysp_2$ & $\lambda(\Ysp_2)$=1/17 & $\lambda_{\rm E}(\Ysp_2)=
\frac{1+d_4}{17-2d_3-d_4}$
&%$\circ$
\\
\cline{3-12}
D0-D8& D8 & $\circ$ & $\circ$ & $\circ$ & $\circ$ & $\circ$ & $\circ$
& $\circ$ & $\circ$ & $\circ$ &&
$\surd$&& &
&%$\circ$
\\
\cline{3-12}
&$x^N$ & $t$ & $v^1$ & $v^2$ & $v^3$ & $v^4$ & $v^5$ & $v^6$
& $v^7$ & $v^8$ & $z$ &
&& &
&%$\circ$
\\
%\cline{1-17}
\hline
\hline
&D1 & $\circ$ &&&   &  &&&& $\circ$ && $\surd$
& $\Ysp_1$ \& $\Zsp$ & $\lambda(\Ysp_1)$=1/5&
$\lambda_{\rm E}(\Ysp_1)=\frac{2-d_3}{10-d_2-2d_3-d_4}$
&%$\circ$
\\
\hline
\end{tabular}
\end{center}
}
\label{table_8}
\end{table}

%\begin{table}[ht]
\begin{table}[p]
\caption{\baselineskip 14pt
Intersections of two D-branes with $p=1$ intersection
in case (III).
%Here is no difference whichever of the two D5's is time dependent.
Whichever of the two D5' is time dependent does not 
make any difference. 
The static D3-D7 solution provides a realistic brane world model. }
\label{twoDIII-2}
{\scriptsize
\begin{center}
\begin{tabular}{|c||c|c|c|c|c|c|c|c|c|c|c||c||c|c|c|c|}
\hline
Branes&&0&1&2&3&4&5&6&7&8&9& & $\tilde{\Msp}$ & $\lambda(\tilde{\Msp})$
& $\lambda_{\rm E}(\tilde{\Msp})$&
BW
\\
\hline
&D2 & $\circ$ & $\circ$ &&   &  &&&&& $\circ$ & $\surd$
& $\Ysp_1$ & $\lambda(\Ysp_1)$=3/11 &
$\lambda_{\rm E}(\Ysp_1)=\frac{-3+d_1+d_3}{-11+2d_1+d_2+2d_3}$
&%$\circ$
\\
\cline{3-12}
D2-D8&D8 & $\circ$ & $\circ$ & $\circ$ & $\circ$ &$\circ$& $\circ$ & $\circ$ &
 $\circ$ & $\circ$ &&& & &
&%$\circ$
\\
\cline{3-12}
&$x^N$ & $t$ & $x$ & $y^1$ & $y^2$ & $y^3$ & $y^4$ & $y^5$
& $y^6$ & $y^7$ & $v$ &
&& &
&%$\circ$
\\
%\cline{2-17}
\hline
&D2 & $\circ$ & $\circ$ & $\circ$ &   &   &&&&&&
& $\Ysp_2$ & $\lambda(\Ysp_2)=1/17$ &
$\lambda_{\rm E}(\Ysp_2)=\frac{-1-d_2}{-17+2d_1+d_2+2d_3}$
&%$\circ$
\\
\cline{3-12}
D2-D8&D8 & $\circ$ & $\circ$ && $\circ$ & $\circ$& $\circ$ & $\circ$ &
 $\circ$ & $\circ$ & $\circ$ &
 $\surd$ & & &
&%$\circ$
\\
\cline{3-12}
&$x^N$ & $t$ & $x$ & $y$ & $v^1$ & $v^2$ & $v^3$ & $v^4$
& $v^5$ & $v^6$ & $v^7$ &
&& &
&%$\circ$
\\
\hline
\hline
&D3 & $\circ$ & $\circ$ &&&  &&&& $\circ$ & $\circ$ & $\surd$
& $\tilde{\Xsp}$ \& $\Ysp_2$ & $\lambda(\tilde{\Xsp})=-1/3$ &
$\lambda_{\rm E}(\tilde{\Xsp})=\frac{d_2-4}{12-2d_1-d_2-2d_3}$
&$(({\cal \Xsp}, \Ysp_2), \Ysp_1)$, I(s)
%$\circ$
\\
\hline
\end{tabular}
\end{center}
}
\label{table_9}
\end{table}

%\begin{table}[ht]
\begin{table}[p]
\caption{\baselineskip 14pt
Intersections of F1-branes in cases (I) and (II).
The F1-D3 solution can provide a realistic brane world model.
}
\label{twoFI}
{\scriptsize
\begin{center}
\begin{tabular}{|c||c|c|c|c|c|c|c|c|c|c|c||c||c|c|c|c|}
\hline
Branes&&0&1&2&3&4&5&6&7&8&9& & $\tilde{\Msp}$ & $\lambda(\tilde{\Msp})$
& $\lambda_{\rm E}(\tilde{\Msp})$&
BW
\\
\hline
&F1 & $\circ$ & $\circ$ &&   &  &&&&&& $\surd$
& $\Ysp_1$ \& $\Zsp$ & $\lambda(\Ysp_1)$=1/5&
$\lambda_{\rm E}(\Ysp_1)=\frac{2-d_1}{10-2d_1-d_2-d_4}$
&%$\circ$
\\
\cline{3-12}
F1-NS5&NS5 & $\circ$ & $\circ$ & $\circ$ & $\circ$ &$\circ$& $\circ$
& & && &
& &  $\lambda(\Zsp)$=1/5&
$\lambda_{\rm E}(\Zsp)=\frac{2-d_1}{10-2d_1-d_2-d_4}$
&%$\circ$
\\
\cline{3-12}
&$x^N$ & $t$ & $x$ & $y^1$ & $y^2$ & $y^3$ & $y^4$ & $z^1$
& $z^2$ & $z^3$ & $z^4$ &
&& &
&%$\circ$
\\
%\cline{2-17}
\hline
&F1 & $\circ$ & $\circ$ & &   &   &&&&&&
& $\tilde{\Xsp}$ \& $\Ysp_2$ & $\lambda(\tilde{\Xsp})=-1/7$ &
$\lambda_{\rm E}(\tilde{\Xsp})=\frac{-2+d_4}{14-2d_1-2d_3-d_4}$
&%$\circ$
\\
\cline{3-12}
F1-NS5&NS5 & $\circ$ & $\circ$ & $\circ$ & $\circ$ & $\circ$& $\circ$
 & & & & &
 $\surd$ & or $\Zsp$ & $\lambda(\Ysp_2)=-1/7$ &
$\lambda_{\rm E}(\Ysp_2)=\frac{-2+d_4}{14-2d_1-2d_3-d_4}$
&%$\circ$
\\
\cline{3-12}
&$x^N$ & $t$ & $x$ & $v^1$ & $v^2$ & $v^3$ & $v^4$ & $z^1$
& $z^2$ & $z^3$ & $z^4$ &
&&  $\lambda(\Zsp)=3/7$ &
$\lambda_{\rm E}(\Zsp)=\frac{6-d_1-d_3}{14-2d_1-2d_3-d_4}$
&%$\circ$
\\
\hline
\hline
&F1 & $\circ$ && $\circ$ &&  &&&&&& $\surd$
&  $\Zsp$ & $\lambda(\Zsp)=1/5$&
$\lambda_{\rm E}(\Zsp)=\frac{2-d_3}{10-d_2-2d_3-d_4}$
&%$\circ$
\\
\hline
\end{tabular}
\end{center}
}
\label{table_10}
\end{table}

%\begin{table}[ht]
\begin{table}[p]
\caption{\baselineskip 14pt
Intersections of F1-branes
in cases (I) and (II).
No solution provides a realistic brane world model.
}
\label{twoFI-2}
{\scriptsize
\begin{center}
\begin{tabular}{|c||c|c|c|c|c|c|c|c|c|c|c||c||c|c|c|c|}
\hline
Branes&&0&1&2&3&4&5&6&7&8&9& & $\tilde{\Msp}$ & $\lambda(\tilde{\Msp})$
& $\lambda_{\rm E}(\tilde{\Msp})$&
BW
\\
\hline
&F1 & $\circ$ &&&   &  &&& $\circ$ &&& $\surd$
& $\Ysp_1$ \& $\Zsp$ & $\lambda(\Ysp_1)$=1/5&
$\lambda_{\rm E}(\Ysp_1)=\frac{2-d_3}{10-d_2-2d_3-d_4}$
&%$\circ$
\\
\cline{3-12}
F1-D6&D6 & $\circ$ & $\circ$ & $\circ$ & $\circ$ &$\circ$& $\circ$ & $\circ$ &
 & & && & $\lambda(\Zsp)$=1/5 &
 $\lambda_{\rm E}(\Zsp)=\frac{2-d_3}{10-d_2-2d_3-d_4}$
&%$\circ$
\\
\cline{3-12}
&$x^N$ & $t$ & $y^1$ & $y^2$ & $y^3$ & $y^4$ & $y^5$ & $y^6$
& $v$ & $z^1$ & $z^2$ &
&& &
&%$\circ$
\\
%\cline{2-17}
\hline
&F1 & $\circ$ & $\circ$ & &   &   &&&&&&
& $\Ysp_2$ or & $\lambda(\Ysp_2)=-1/15$ &
$\lambda_{\rm E}(\Ysp_2)=\frac{d_2+d_4-1}{15-d_2-2d_3-d_4}$
&%$\circ$
\\
\cline{3-12}
F1-D6&D6 & $\circ$ & & $\circ$ & $\circ$ & $\circ$& $\circ$ &  $\circ$ &
  $\circ$  && &
 $\surd$ & $\Ysp_1$ \& $\Zsp$ & $\lambda(\Ysp_1)=7/15$ &
 $\lambda_{\rm E}(\Ysp_1)=\frac{7-d_3}{15-d_2-2d_3-d_4}$
&%$\circ$
\\
\cline{3-12}
&$x^N$ & $t$ & $y$ & $v^1$ & $v^2$ & $v^3$ & $v^4$ & $v^5$
& $v^6$ & $z^1$ & $z^2$ &
&& $\lambda(\Zsp)=7/15$ & $\lambda_{\rm E}(\Zsp)=\frac{7-d_3}{15-d_2-2d_3-d_4}$
&%$\circ$
\\
\hline
\hline
&F1 & $\circ$ &&&   &  &&&& $\circ$ && $\surd$
& $\Ysp_1$ \& $\Zsp$ & $\lambda(\Ysp_1)$=1/5&
$\lambda_{\rm E}(\Ysp_1)=\frac{2-d_3}{10-d_2-2d_3-d_4}$
&%$\circ$
\\
\hline
\end{tabular}
\end{center}
}
\label{table_11}
\end{table}

%\begin{table}[ht]
\begin{table}[p]
\caption{\baselineskip 14pt
Intersections of NS-branes
in cases (I) and (II).
%Here is no difference whichever of
%the two NS5's is time dependent.
Whichever of the two NS5' is time dependent does not 
make any difference. 
D3-NS5 and D4-NS5 solutions can provide realistic brane world models.
}
\label{twoNSI}
{\scriptsize
\begin{center}
\begin{tabular}{|c||c|c|c|c|c|c|c|c|c|c|c||c||c|c|c|c|}
\hline
Branes&&0&1&2&3&4&5&6&7&8&9& & $\tilde{\Msp}$ & $\lambda(\tilde{\Msp})$
& $\lambda_{\rm E}(\tilde{\Msp})$
& BW
\\
\hline
&NS5 & $\circ$ & $\circ$ & $\circ$ & $\circ$ &&& $\circ$ & $\circ$ &&& $\surd$
& $\tilde{\Xsp}$ \& $\Ysp_2$ & $\lambda(\tilde{\Xsp})=\lambda(\Ysp_2)=
\frac{-1}{7}$&
$\lambda_{\rm E}(\tilde{\Xsp})=\lambda_{\rm E}(\Ysp_2)=
\frac{d_2+d_4-2}{14-2d_1-d_2-2d_3-d_4}$
&%$\circ$
\\
\cline{3-12}
NS5&NS5 & $\circ$ &$\circ$&$\circ$ & $\circ$ & $\circ$ & $\circ$ &&
&&& & or & $\lambda(\Ysp_1)=
 \lambda(\Zsp)=\frac{3}{7}$ &
$\lambda_{\rm E}(\Ysp_1)=\lambda_{\rm E}(\Zsp)=
\frac{6-d_1-d_3}{14-2d_1-d_2-2d_3-d_4}$
&%$\circ$
\\
\cline{3-12}
&$x^N$ & $t$ & $x^1$ & $x^2$ & $x^3$ & $y^1$ & $y^2$ & $v^1$
& $v^2$ & $z^1$ & $z^2$ &
&$\Ysp_1$ \& $\Zsp$ &  &
&%$\circ$
\\
%\cline{2-17}
\hline
\hline
&D1 & $\circ$ &&&   &  && $\circ$ &&&& $\surd$
& $\Ysp_1$ \& $\Zsp$ & $\lambda(\Ysp_1)$=1/5&
$\lambda_{\rm E}(\Ysp_1)=\frac{2-d_3}{10-d_2-2d_3-d_4}$
&%$\circ$
\\
\cline{3-12}
D1-NS5&NS5 & $\circ$ & $\circ$ & $\circ$ & $\circ$ &$\circ$& $\circ$
& & && &
& &  $\lambda(\Zsp)$=1/5&
$\lambda_{\rm E}(\Zsp)=\frac{2-d_3}{10-d_2-2d_3-d_4}$
&%$\circ$
\\
\cline{3-12}
&$x^N$ & $t$ & $y^1$ & $y^2$ & $y^3$ & $y^4$ & $y^5$ & $v$
& $z^1$ & $z^2$ & $z^3$ &
&& &
&%$\circ$
\\
%\cline{2-17}
\hline
&D1 & $\circ$ & $\circ$ & &   &   &&&&&&
& $\Ysp_2$ & $\lambda(\Ysp_2)=-1/7$ &
$\lambda_{\rm E}(\Ysp_2)=\frac{d_2+d_4-2}{14-d_2-2d_3-d_4}$
&%$\circ$
\\
\hline
\end{tabular}
\end{center}
}
\label{table_12}
\end{table}

%\begin{table}[ht]
\begin{table}[p]
\caption{\baselineskip 14pt
Intersections of NS-branes in case (III).
%Here is no difference whichever of
%the two NS5's is time dependent.
Whichever of the two NS5' is time dependent does not 
make any difference. 
D7-NS5 solutions can provide realistic brane world models.
}
\label{twoNSIII}
{\scriptsize
\begin{center}
\begin{tabular}{|c||c|c|c|c|c|c|c|c|c|c|c||c||c|c|c|c|}
\hline
Branes&&0&1&2&3&4&5&6&7&8&9& & $\tilde{\Msp}$ & $\lambda(\tilde{\Msp})$
& $\lambda_{\rm E}(\tilde{\Msp})$&
BW
\\
\hline
&NS5 & $\circ$ & $\circ$ &&&&& $\circ$ & $\circ$ & $\circ$ & $\circ$ & $\surd$
& $\tilde{\Xsp}$ \& $\Ysp_2$ & $\lambda(\tilde{\Xsp})=-1/7$&
$\lambda_{\rm E}(\tilde{\Xsp})=\frac{d_2-2}{14-2d_1-d_2-2d_3}$
&%$\circ$
\\
\cline{3-12}
NS5&NS5 & $\circ$ &$\circ$& $\circ$ & $\circ$ & $\circ$ & $\circ$ &&
&&& & or $\Ysp_1$ & $\lambda(\Ysp_2)=-1/7$ &
$\lambda_{\rm E}(\Ysp_2)=\frac{d_2-2}{14-2d_1-d_2-2d_3}$
&%$\circ$
\\
\cline{3-12}
&$x^N$ & $t$ & $x$ & $y^1$ & $y^2$ & $y^3$ & $y^4$ & $v^1$
& $v^2$ & $v^3$ & $v^4$ &
&& $\lambda(\Ysp_1)$=3/7 &
$\lambda_{\rm E}(\Ysp_1)=\frac{6-d_1-d_3}{14-2d_1-d_2-2d_3}$
&%$\circ$
\\
%\cline{2-17}
\hline
\hline
&D3 & $\circ$ &&&&  && $\circ$ & $\circ$ & $\circ$ && $\surd$
&  $\Ysp_2$ or & $\lambda(\Ysp_2)=-1/3$&
$\lambda_{\rm E}(\Ysp_2)=\frac{d_2+d_4-4}{12-d_2-2d_3-d_4}$
&
%$\circ$
\\
\cline{3-12}
D3-NS5&NS5 & $\circ$ & $\circ$ & $\circ$ & $\circ$ & $\circ$& $\circ$ &&
&&& & $\Ysp_1$ \& $\Zsp$ & $\lambda(\Ysp_1)$=1/3 &
$\lambda_{\rm E}(\Ysp_1)=\frac{4-d_3}{12-d_2-2d_3-d_4}$
&%$\circ$
\\
\cline{3-12}
&$x^N$ & $t$ & $y^1$ & $y^2$ & $y^3$ & $y^4$ & $y^5$ & $v^1$
& $v^2$ & $v^3$ & $z$ &
&&$\lambda(\Zsp)$=1/3 &
$\lambda_{\rm E}(\Zsp)=\frac{4-d_3}{12-d_2-2d_3-d_4}$
&%$\circ$
\\
%\cline{2-17}
\hline
&D3 & $\circ$ & $\circ$ & $\circ$ & $\circ$ &   &&&&&&
& $\Ysp_2$  or & $\lambda(\Ysp_2)=-1/7$&
$\lambda_{\rm E}(\Ysp_2)=\frac{d_2+d_4-2}{14-d_2-2d_3-d_4}$
&%$\circ$
\\
\hline
\end{tabular}
\end{center}
}
\label{table_13}
\end{table}

\begin{table}[p]
\caption{\baselineskip 14pt
The power exponent of the fastest expansion in the Einstein frame
for M-brane. ``TD" in the table shows which brane is time dependent.
%Only the brane systems marked
%in the column ``BW" have brane world ``BW" spacetime.
}
\begin{center}
{\scriptsize
\begin{tabular}{|c|c|c||c|c||c||c|}
\hline
Branes& TD &dim$(\Msp)$  &$\tilde{\Msp}$
&$(d_1, d_2, d_3, d_4)$&
$\lambda_{\rm E}(\tilde{\Msp})$ & Case
\\
\hline
\hline
M2-M2&M2& 9& $\Ysp_1$ \& Z & (0, 2, 0, 0) &3/10& I
\\
\cline{2-7}
&M2&9& $\Ysp_1$ \& $\Ysp_2$ \& Z & (0, 0, 1, 0) & 1/5 & II
\\
\hline
&M2&7& $\tilde{\Xsp}$ \& $\Ysp_2$ \& Z & (0, 4, 0, 0)  & 3/8 & I
\\
\cline{2-7}
&M2&10& $\Ysp_1$ \& $\Ysp_2$ \& Z & (1, 0, 0, 0) &1/5 & II
\\
\cline{2-7}
M2-M5&M2&10& $\tilde{\Xsp}$ \& $\Ysp_1$ \& Z & (0, 0, 1, 0) &1/5 & II
\\
\cline{2-7}
&M5&10& $\tilde{\Xsp}$ \& $\Ysp_2$ \& Z
& (0, 1, 0, 0) &3/7 & I
\\
\cline{2-7}
&M5&10& $\Ysp_1$ \& $\Ysp_2$ \& Z & (1, 0, 0, 0) &5/13 & II
\\
\cline{2-7}
&M5&10& $\tilde{\Xsp}$ \& $\Ysp_1$ \& $\Ysp_2$ \& Z &
(0, 0, 1, 0) &5/13 & II
\\
\hline
&M5&9& $\tilde{\Xsp}$ \& $\Ysp_2$ \& Z & (0, 2, 0, 0)& 6/13 &  I
\\
%\hline
\cline{2-7}
M5-M5&M5&10&$\tilde{\Xsp}$ \& $\Ysp_1$ \& $\Ysp_2$ \& Z
& (1, 0, 0, 0) & 5/13 & II
\\
\cline{2-7}
&M5&10&$\tilde{\Xsp}$ \& $\Ysp_1$ \& $\Ysp_2$ \& Z
&(0, 0, 1, 0) & 5/13 & II
\\
\cline{2-7}
&M5&10&$\tilde{\Xsp}$ \& $\Ysp_1$ \& $\Ysp_2$ & (0, 0, 0, 1) & 3/7 & III
\\
\hline
\end{tabular}
}
\label{table_14}
\end{center}
\end{table}

\begin{table}[p]
\caption{\baselineskip 14pt
The power exponent of the fastest expansion in the Einstein frame
for D-brane of cases (I) and (II) ($p=0$).
``TD" in the table shows which brane is time dependent.
%Only the brane systems marked
%in the column ``BW" have brane world ``BW" spacetime.
}
\begin{center}
{\scriptsize
\begin{tabular}{|c|c|c||c|c||c||c|}
\hline
Branes& TD &dim$(\Msp)$  &$\tilde{\Msp}$
&$(d_1, d_2, d_3, d_4)$&
$\lambda_{\rm E}(\tilde{\Msp})$ & Case
\\
\hline
\hline
&D0&6& $\Zsp$  & (0, 4, 0, 0) &1/5& I
\\
\cline{2-7}
D0-D4&D0&10& $\Ysp_1$ \& $\Zsp$  & (0, 0, 0, 0) &1/9& II
\\
\cline{2-7}
&D4&9&$\Ysp_2$ \& $\Zsp$ & (0, 0, 1, 0) &4/11& I \& II
\\
\hline
&D1&7& $\Ysp_2$ \& $\Zsp$ & (0, 3, 0, 0)& 2/7 & I
\\
\cline{2-7}
D1-D3&D1&9& $\Ysp_1$ \& $\Zsp$ & (0, 0, 1, 0) &1/8& II
\\
\cline{2-7}
&D3&9& $\Ysp_2$ \& $\Zsp$ & (0, 1, 0, 0) &4/11& I
\\
\cline{2-7}
&D3&9& $\Ysp_1$ \& $\Ysp_2$ \& $\Zsp$ & (0, 0, 1, 0) &3/10& II
\\
\hline
D2-D2&D2&8& $\Ysp_2$ \& $\Zsp$ & (0, 2, 0, 0)& 1/3 & I
\\
\cline{2-7}
&D2&9& $\Ysp_1$ \& $\Ysp_2$ \& $\Zsp$ & (0, 0, 1, 0) &2/9& II
\\
\hline
\end{tabular}
}
\label{table_15}
\end{center}
\end{table}

\begin{table}[p]
\caption{\baselineskip 14pt
The power exponent of the fastest expansion in the Einstein frame
for D-brane of cases (I) and (II) ($p=1$).
``TD" in the table shows which brane is time dependent.
%Only the brane systems marked
%in the column ``BW" have brane world ``BW" spacetime.
}
\begin{center}
{\scriptsize
\begin{tabular}{|c|c|c||c|c||c||c|}
\hline
Branes& TD &dim$(\Msp)$  &$\tilde{\Msp}$
&$(d_1, d_2, d_3, d_4)$&
$\lambda_{\rm E}(\tilde{\Msp})$ & Case
\\
\hline
\hline
&D1&6& $\tilde{\Xsp}$ \& $\Zsp$  & (0, 4, 0, 0) &1/3& I
\\
\cline{2-7}
D1-D5&D1&9& $\Ysp_1$ \& $\Zsp$  & (1, 0, 0, 0) &1/8& II
\\
\cline{2-7}
&D5&9& $\Ysp_2$ \& $\Zsp$ & (1, 0, 0, 0) &5/12& I \& II
\\
\cline{2-7}
&D5&9& $\tilde{\Xsp}$ \& $\Ysp_2$ \& $\Zsp$ & (0, 0, 1, 0) &5/12& I \& II
\\
\hline
&D2&7& $\tilde{\Xsp}$ \& $\Ysp_2$ \& $\Zsp$ & (0, 3, 0, 0)& 3/8 & I
\\
\cline{2-7}
&D2&9& $\Ysp_1$ \& $\Ysp_2$ \&$\Zsp$ & (1, 0, 0, 0) &2/9& II
\\
\cline{2-7}
D2-D4&D2&9& $\tilde{\Xsp}$ \& $\Ysp_1$ \& $\Zsp$ & (0, 0, 1, 0) &2/9& II
\\
\cline{2-7}
&D4&9& $\tilde{\Xsp}$ \& $\Ysp_2$ \& $\Zsp$ & (0, 1, 0, 0) &5/12& I
\\
\cline{2-7}
&D4&9& $\Ysp_1$ \& $\Ysp_2$ \& $\Zsp$ & (1, 0, 0, 0) &4/11& II
\\
\cline{2-7}
&D4&9& $\tilde{\Xsp}$ \& $\Ysp_1$ \& $\Ysp_2$ \& $\Zsp$ &(0, 0, 1, 0)
&4/11& II
\\
\hline
&D3&8& $\tilde{\Xsp}$ \& $\Ysp_2$ \& $\Zsp$ & (0, 2, 0, 0)& 2/5 & I
\\
\cline{2-7}
D3-D3&D3&9& $\Ysp_1$ \& $\Ysp_2$ \& $\Zsp$ & (1, 0, 0, 0) &3/10& II
\\
\cline{2-7}
&D3&9& $\tilde{\Xsp}$ \& $\Ysp_1$ \& $\Ysp_2$ \& $\Zsp$ & (0, 0, 1, 0) &3/10& II
\\
\hline
\end{tabular}
}
\label{table_16}
\end{center}
\end{table}

\begin{table}[p]
\caption{\baselineskip 14pt
The power exponent of the fastest expansion in the Einstein frame
for D-brane of cases (I) and (II) ($p=2$).
``TD" in the table shows which brane is time dependent.
%Only the brane systems marked
%in the column ``BW" have brane world ``BW" spacetime.
}
\begin{center}
{\scriptsize
\begin{tabular}{|c|c|c||c|c||c||c|}
\hline
Branes& TD &dim$(\Msp)$  &$\tilde{\Msp}$
&$(d_1, d_2, d_3, d_4)$&
$\lambda_{\rm E}(\tilde{\Msp})$ & Case
\\
\hline
\hline
&D2&6& $\tilde{\Xsp}$ \& $\Zsp$  & (0, 4, 0, 0) &3/7& I
\\
\cline{2-7}
D2-D6&D2&9& $\tilde{\Xsp}$ \& $\Ysp_1$ \& $\Zsp$  & (1, 0, 0, 0) &2/9& II
\\
\cline{2-7}
&D6&9& $\tilde{\Xsp}$ \& $\Ysp_2$ \& $\Zsp$ & (1, 0, 0, 0) &6/13& I \& II
\\
\cline{2-7}
&D6&9& $\tilde{\Xsp}$ \& $\Ysp_2$ \& $\Zsp$ & (0, 0, 1, 0) &6/13& I \& II
\\
\hline
&D3&7& $\tilde{\Xsp}$ \& $\Ysp_2$ \& $\Zsp$ & (0, 3, 0, 0)& 4/9 & I
\\
\cline{2-7}
&D3&9& $\tilde{\Xsp}$ \& $\Ysp_1$ \& $\Ysp_2$ \&$\Zsp$
& (1, 0, 0, 0) &3/10& II
\\
\cline{2-7}
D3-D5&D3&9& $\tilde{\Xsp}$ \& $\Ysp_1$ \& $\Zsp$
& (0, 0, 1, 0) &3/10& II
\\
\cline{2-7}
&D5&9& $\tilde{\Xsp}$ \& $\Ysp_2$ \& $\Zsp$ & (0, 1, 0, 0) &6/13& I
\\
\cline{2-7}
&D5&9& $\tilde{\Xsp}$ \& $\Ysp_1$ \& $\Ysp_2$ \& $\Zsp$
& (1, 0, 0, 0) &5/12& II
\\
\cline{2-7}
&D5&9&$\tilde{\Xsp}$ \& $\Ysp_1$ \& $\Ysp_2$ \& $\Zsp$
& (0, 0, 1, 0) &5/12& II
\\
\hline
&D4&8& $\tilde{\Xsp}$ \& $\Ysp_2$ \& $\Zsp$ & (0, 2, 0, 0)& 5/11 & I
\\
\cline{2-7}
D4-D4&D4&9& $\tilde{\Xsp}$ \& $\Ysp_1$ \& $\Ysp_2$ \& $\Zsp$
& (1, 0, 0, 0) &4/11& II
\\
\cline{2-7}
&D4&9&$\tilde{\Xsp}$ \& $\Ysp_1$ \& $\Ysp_2$ \& $\Zsp$ & (0, 0, 1, 0)
&4/11& II
\\
\hline
\end{tabular}
}
\label{table_17}
\end{center}
\end{table}

\begin{table}[p]
\caption{\baselineskip 14pt
The power exponent of the fastest expansion in the Einstein frame
for D-brane of cases (I) and (II) ($p=3$).
``TD" in the table shows which brane is time dependent.
%Only the brane systems marked
%in the column ``BW" have brane world ``BW" spacetime.
}
\begin{center}
{\scriptsize
\begin{tabular}{|c|c|c||c|c||c||c|}
\hline
Branes& TD &dim$(\Msp)$  &$\tilde{\Msp}$
&$(d_1, d_2, d_3, d_4)$&
$\lambda_{\rm E}(\tilde{\Msp})$ & Case
\\
\hline
\hline
&D3&7& $\tilde{\Xsp}$ \& $\Ysp_1$ \& $\Zsp$  & (0, 3, 0, 0) &4/9& I
\\
\cline{2-7}
D3-D7&D3&9& $\tilde{\Xsp}$ \& $\Ysp_1$ \& $\Zsp$  & (1, 0, 0, 0) &3/10& II
\\
\cline{2-7}
&D7&$10-d$& $\tilde{\Xsp}$ \& $\Ysp_2$ \& $\Zsp$ or $\tilde{\Xsp}$ \& $\Zsp$
or $\Ysp_2$ \& $\Zsp$ & ($d_1$, 0, $d_3$, 0)  &0& I \& II
\\
\hline
&D4&8& $\tilde{\Xsp}$ \& $\Ysp_1$ \& $\Ysp_2$ \& $\Zsp$ & (0, 2, 0, 0)
& 5/11 & I
\\
\cline{2-7}
&D4&9& $\tilde{\Xsp}$ \& $\Ysp_1$ \& $\Ysp_2$ \&$\Zsp$
& (1, 0, 0, 0) &4/11& II
\\
\cline{2-7}
D4-D6&D4&9& $\tilde{\Xsp}$ \& $\Ysp_1$ \& $\Zsp$ & (0, 0, 1, 0) &4/11& II
\\
\cline{2-7}
&D6&9& $\tilde{\Xsp}$ \& $\Ysp_1$ \& $\Ysp_2$ \& $\Zsp$
& (1, 0, 0, 0) &6/13& I \& II
\\
\cline{2-7}
&D6&9& $\tilde{\Xsp}$ \& $\Ysp_1$ \& $\Ysp_2$ \& $\Zsp$
& (0, 0, 1, 0) &6/13& I \& II
\\
\hline
&D5&9& $\tilde{\Xsp}$ \& $\Ysp_1$ \& $\Ysp_2$ \& $\Zsp$
& (0, 1, 0, 0)& 6/13 & I
\\
\cline{2-7}
D5-D5&D5&9& $\tilde{\Xsp}$ \& $\Ysp_1$ \& $\Ysp_2$ \& $\Zsp$
& (1, 0, 0, 0) &5/12& II
\\
\cline{2-7}
&D5&9& $\tilde{\Xsp}$ \& $\Ysp_1$ \& $\Ysp_2$ \& $\Zsp$
&(0, 0, 1, 0) &5/12& II
\\
\hline
\end{tabular}
}
\label{table_18}
\end{center}
\end{table}

\begin{table}[p]
\caption{\baselineskip 14pt
The power exponent of the fastest expansion in the Einstein frame
for D-brane of cases (I) and (II) ($p=4$).
``TD" in the table shows which brane is time dependent.
%Only the brane systems marked
%in the column ``BW" have brane world ``BW" spacetime.
}
\begin{center}
{\scriptsize
\begin{tabular}{|c|c|c||c|c||c||c|}
\hline
Branes& TD &dim$(\Msp)$  &$\tilde{\Msp}$
&$(d_1, d_2, d_3, d_4)$&
$\lambda_{\rm E}(\tilde{\Msp})$ & Case
\\
\hline
\hline
&D4&8& $\tilde{\Xsp}$ \& $\Ysp_1$ \& $\Zsp$  & (0, 2, 0, 0) &5/11 & I
\\
\cline{2-7}
&D4&9& $\tilde{\Xsp}$ \& $\Ysp_1$ \& $\Zsp$  & (1, 0, 0, 0) &4/11& II
\\
\cline{2-7}
D4-D8&D8&5& $\Ysp_2$ \& $\Zsp$ & (4, 0, 1, 0) &1/7& I \& II
\\
\cline{2-7}
&D8&5&$\tilde{\Xsp}$ \& $\Zsp$ & (1, 0, 4, 0) &1/7& I \& II
\\
\cline{2-7}
&D8&5& $\tilde{\Xsp}$ \& $\Ysp_2$ \& $\Zsp$ & (3, 0, 2, 0) &1/7& I \& II
\\
\cline{2-7}
&D8&5& $\tilde{\Xsp}$ \& $\Ysp_2$ \& $\Zsp$ & (2, 0, 3, 0) &1/7& I \& II
\\
\hline
&D5&9& $\tilde{\Xsp}$ \& $\Ysp_1$ \& $\Ysp_2$ \& $\Zsp$
& (0, 1, 0, 0)& 6/13 & I
\\
\cline{2-7}
&D5&9& $\tilde{\Xsp}$ \& $\Ysp_1$ \& $\Ysp_2$ \& $\Zsp$
& (1, 0, 0, 0)  &5/12& II
\\
\cline{2-7}
&D5&9& $\tilde{\Xsp}$ \& $\Ysp_2$ \& $\Zsp$
& (0, 0, 1, 0) &5/12& II
\\
\cline{2-7}
D5-D7&D7&5& $\Ysp_2$ \& $\Zsp$ & (4, 1, 0, 0) &1/6& I
\\
\cline{2-7}
&D7&5& $\tilde{\Xsp}$ \& $\Ysp_2$ \& $\Zsp$ & (3, 1, 1, 0) &1/6& I
\\
\cline{2-7}
&D7&5& $\tilde{\Xsp}$ \& $\Ysp_2$ \& $\Zsp$ & (2, 1, 2, 0) &1/6& I
\\
\cline{2-7}
&D7&5& $\tilde{\Xsp}$ \& $\Zsp$ & (1, 1, 3, 0) &1/6& I
\\
\cline{2-7}
&D7&$10-d$ & $\tilde{\Xsp}$ \& $\Ysp_1$ \& $\Ysp_2$ \& $\Zsp$
or $\tilde{\Xsp}$ \& $\Ysp_1$ \& $\Zsp$
or $\Ysp_1$ \& $\Ysp_2$ \& $\Zsp$
& ($d_1$, 0, $d_3$, 0) &0& II
\\
\hline
D6-D6&D6&9& $\tilde{\Xsp}$ \& $\Ysp_1$
\& $\Ysp_2$ \& $\Zsp$ & (1, 0, 0, 0) &6/13& I \& II
\\
\cline{2-7}
&D6&9& $\tilde{\Xsp}$ \& $\Ysp_1$
\& $\Ysp_2$ \& $\Zsp$ & (0, 0, 1, 0) &6/13& I \& II
\\
\hline
\end{tabular}
}
\label{table_19}
\end{center}
\end{table}

\begin{table}[p]
\caption{\baselineskip 14pt
The power exponent of the fastest expansion in the Einstein frame
for D-brane of cases (I) and (II) ($p=5$).
``TD" in the table shows which brane is time dependent.
%Only the brane systems marked
%in the column ``BW" have brane world ``BW" spacetime.
}
\begin{center}
{\scriptsize
\begin{tabular}{|c|c|c||c|c||c||c|}
\hline
Branes& TD &dim$(\Msp)$  &$\tilde{\Msp}$
&$(d_1, d_2, d_3, d_4)$&
$\lambda_{\rm E}(\tilde{\Msp})$ & Case
\\
\hline
\hline
&D6&9& $\tilde{\Xsp}$ \& $\Ysp_1$ \& $\Ysp_2$ \& $\Zsp$
& (1, 0, 0, 0) &6/13& I \& II
\\
\cline{2-7}
&D6&9& $\tilde{\Xsp}$ \& $\Ysp_1$ \& $\Ysp_2$ \& $\Zsp$
&  (0, 0, 1, 0) &6/13& I \& II
\\
\cline{2-7}
&D8&5& $\Ysp_1$ \& $\Ysp_2$ & (5, 1, 0, 0)& 1/3 & I
\\
\cline{2-7}
&D8&5& $\tilde{\Xsp}$ \& $\Ysp_2$ & (4, 1, 1, 0) &1/3& I
\\
\cline{2-7}
&D8&5& $\tilde{\Xsp}$ \& $\Ysp_2$ & (3, 1, 2, 0) &1/3& I
\\
\cline{2-7}
D6-D8&D8&5& $\tilde{\Xsp}$ & (2, 1, 3, 0)& 1/3 & I
\\
\cline{2-7}
&D8&6& $\Ysp_1$ \& $\Ysp_2$ & (5, 0, 0, 0) &1/7& II
\\
\cline{2-7}
&D8&6& $\tilde{\Xsp}$ \& $\Ysp_1$ \& $\Ysp_2$ & (4, 0, 1, 0) &1/7& II
\\
\cline{2-7}
&D8&6& $\tilde{\Xsp}$ \& $\Ysp_1$ \& $\Ysp_2$ & (3, 0, 2, 0) &1/7& II
\\
\cline{2-7}
&D8&6& $\tilde{\Xsp}$ \& $\Ysp_1$ \& $\Ysp_2$ & (2, 0, 3, 0) &1/7& II
\\
\hline
&D7&4& $\tilde{\Xsp}$ \& $\Ysp_2$ & (4, 2, 0, 0) & 1/3 & I
\\
\cline{2-7}
&D7&4& $\tilde{\Xsp}$ \& $\Ysp_2$ &(3, 2, 1, 0) & 1/3 & I
\\
\cline{2-7}
D7-D7&D7&4& $\tilde{\Xsp}$ & (2, 2, 2, 0) & 1/3 & I
\\
\cline{2-7}
&D7&$10-d$& $\tilde{\Xsp}$ \& $\Ysp_1$ \& $\Ysp_2$ or
 $\tilde{\Xsp}$ \& $\Ysp_1$
& ($d_1$, 0, $d_3$, 0) &0& II
\\
\hline
\end{tabular}
}
\label{table_20}
\end{center}
\end{table}
%%%%%%%%%%%%%%%%%%%%%%%%%%%%%%%%%%%%%%%%%%%%%%
%%%%%%%%%%%%%%%%%%%%%%%%%%%%%%%%%%%%%%%%%%%%%%
\begin{table}[p]
\caption{\baselineskip 14pt
The power exponent of the fastest expansion in the Einstein frame
for D-brane of case (III) ($p=0$).
``TD" in the table shows which brane is time dependent.
%Only the brane systems marked
%in the column ``BW" have brane world ``BW" spacetime.
}
\begin{center}
{\scriptsize
\begin{tabular}{|c|c|c||c|c||c||c|}
\hline
Branes& TD &dim$(\Msp)$  &$\tilde{\Msp}$
&$(d_1, d_2, d_3, d_4)$&
$\lambda_{\rm E}(\tilde{\Msp})$ & Case
\\
\hline
\hline
D0-D8&D0& 9& $\Ysp_1$
& (0, 0, 0, 1) &1/8& III
\\
\cline{2-7}
&D8&9&$\Ysp_2$ & (0, 0, 0, 1) &1/8& III
\\
\hline
D1-D7&D1&9& $\Ysp_1$ \& $\Ysp_2$ & (0, 0, 0, 1)& 2/9 & III
\\
%\hline
\cline{2-7}
&D7&9& $\Ysp_1$ \& $\Ysp_2$ & (0, 0, 0, 1) &1/15& III
\\
\hline
D2-D6&D2&9& $\Ysp_1$ \& $\Ysp_2$ & (0, 0, 0, 1)& 3/10 & III
\\
%\hline
\cline{2-7}
&D6&9& $\Ysp_1$ \& $\Ysp_2$ \& $\Zsp$& (0, 0, 1, 0) &6/13& III
\\
\hline
D3-D5&D3&9& $\Ysp_1$ \& $\Ysp_2$ & (0, 0, 0, 1)& 4/11 & III
\\
%\hline
\cline{2-7}
&D5&9& $\Ysp_1$ \& $\Ysp_2$ & (0, 0, 0, 1) &6/13& III
\\
\hline
D4-D4&D4&9& $\Ysp_1$ \& $\Ysp_2$ & (0, 0, 0, 1) &5/12& III
\\
\hline
\end{tabular}
}
\label{table_21}
\end{center}
\end{table}

\begin{table}[p]
\caption{\baselineskip 14pt
The power exponent of the fastest expansion in the Einstein frame
for D-brane of case (III) ($p=1$).
``TD" in the table shows which brane is time
dependent.
%Only the brane systems marked
%in the column ``BW" have brane world ``BW" spacetime.
}
\begin{center}
{\scriptsize
\begin{tabular}{|c|c|c||c|c||c||c|}
\hline
Branes& TD &dim$(\Msp)$  &$\tilde{\Msp}$
&$(d_1, d_2, d_3, d_4)$&
$\lambda_{\rm E}(\tilde{\Msp})$ & Case
\\
\hline
\hline
D2-D8&D2& 9& $\Ysp_1$  \& $\Ysp_2$
& (1, 0, 0, 0) &2/9& III
\\
\cline{2-7}
&D8&9&$\Ysp_1$ \& $\Ysp_2$ & (1, 0, 0, 0) &1/15& III
\\
\hline
&D3&9& $\Ysp_1$ \& $\Ysp_2$ & (1, 0, 0, 0)& 3/10 & III
\\
%\hline
\cline{2-7}
D3-D7&D7&$10$& $\tilde{\Xsp}$ \& $\Ysp_1$ \& $\Ysp_2$
& (0, 0, 0, 0) &0& III
\\
\cline{2-7}
&D7&$9$& $\Ysp_1$ \& $\Ysp_2$ & (1, 0, 0, 0) &0& III
\\
\hline
D4-D6&D4&9& $\Ysp_1$ \& $\Ysp_2$ & (1, 0, 0, 0)& 4/11 & III
\\
%\hline
\cline{2-7}
&D6&9& $\Ysp_1$ \& $\Ysp_2$ & (1, 0, 0, 0) &6/13& III
\\
\hline
D5-D5&D5&9& $\Ysp_1$ \& $\Ysp_2$ & (1, 0, 0, 0) &5/12& III
\\
\hline
\end{tabular}
}
\label{table_22}
\end{center}
\end{table}

\begin{table}[p]
\caption{\baselineskip 14pt
The power exponent of the fastest expansion in the Einstein frame
for F1-string of cases (I) and (II). ``TD" in the table shows which brane is time
dependent.
%Only the brane systems marked
%in the column ``BW" have brane world ``BW" spacetime.
}
\begin{center}
{\scriptsize
\begin{tabular}{|c|c|c||c|c||c||c|}
\hline
Branes& TD &dim$(\Msp)$  &$\tilde{\Msp}$
&$(d_1, d_2, d_3, d_4)$& $\lambda_{\rm E}(\tilde{\Msp})$ & Case
\\
\hline
\hline
&F1& 6& $\tilde{\Xsp}$ \& $\Zsp$
& (0, 4, 0, 0) &1/3& I
\\
\cline{2-7}
F1-NS5&F1&9&$\Ysp_1$ \& $\Zsp$ & (1, 0, 0, 0) &1/8& II
\\
\cline{2-7}
&NS5&9&$\Ysp_1$ \& $\Zsp$ & (1, 0, 0, 0) &5/12& I \& II
\\
\cline{2-7}
&NS5&9& $\tilde{\Xsp}$ \& $\Ysp_1$ \& $\Zsp$ & (0, 0, 1, 0) &5/12& I \& II
\\
\hline
&F1&9& $\Ysp_2$ \& $\Zsp$ & (0, 1, 0, 0)& 2/9 & I
\\
\cline{2-7}
F1-D1&F1&9&$\Ysp_1$ \& $\Zsp$ & (0, 0, 1, 0) &1/8& II
\\
\cline{2-7}
&D1&9& $\Ysp_2$ \& $\Zsp$ & (0, 1, 0, 0) &2/9&
I
\\
\hline
\end{tabular}
}
\label{table_23}
\end{center}
\end{table}

\begin{table}[p]
\caption{\baselineskip 14pt
The power exponent of the fastest expansion in the Einstein frame
for NS-brane of cases (I) and (II).
``TD" in the table shows which brane is time dependent.
%Only the brane systems marked
%in the column ``BW" have brane world ``BW" spacetime.
}
\begin{center}
{\scriptsize
\begin{tabular}{|c|c|c||c|c||c||c|}
\hline
Branes& TD &dim$(\Msp)$  &$\tilde{\Msp}$
&$(d_1, d_2, d_3, d_4)$&
$\lambda_{\rm E}(\tilde{\Msp})$ & Case
\\
\hline
\hline
&NS5& 9& $\tilde{\Xsp}$ \& $\Ysp_1$ \& $\Ysp_2$ \& $\Zsp$
& (0, 1, 0, 0) &6/13& I
\\
\cline{2-7}
NS5-NS5&NS5&9& $\tilde{\Xsp}$ \& $\Ysp_1$ \& $\Ysp_2$ \& $\Zsp$
& (1, 0, 0, 0) &5/12& II
\\
\cline{2-7}
&NS5&9& $\tilde{\Xsp}$ \& $\Ysp_1$ \& $\Ysp_2$ \& $\Zsp$
& (0, 0, 1, 0)&5/12& II
\\
\hline
&D1&5& $\Ysp_2$ \& Z & (0, 5, 0, 0)& 2/5 &  I
\\
\cline{2-7}
D1-NS5&D1&9& $\Ysp_1$ \& $\Zsp$ & (0, 0, 1, 0)&1/8& II
\\
\cline{2-7}
&NS5&9& $\Ysp_2$ \& $\Zsp$ & (0, 1, 0, 0)&6/13&
I
\\
\cline{2-7}
&NS5&9&$\Ysp_1$ \& $\Ysp_2$ \& $\Zsp$ & (0, 0, 1, 0)&5/12&
II
\\
\hline
\end{tabular}
}
\label{table_24}
\end{center}
\end{table}

\begin{table}[p]
\caption{\baselineskip 14pt
The power exponent of the fastest expansion in the Einstein frame
for NS-brane of case (III). ``TD" in the table shows which brane is time dependent.
%Only the brane systems marked
%in the column ``BW" have brane world ``BW" spacetime.
}
\begin{center}
{\scriptsize
\begin{tabular}{|c|c|c||c|c||c||c|}
\hline
Branes& TD &dim$(\Msp)$  &$\tilde{\Msp}$
&$(d_1, d_2, d_3, d_4)$&
$\lambda_{\rm E}(\tilde{\Msp})$ & Case
\\
\hline
\hline
NS5-NS5&NS5& 9& $\Ysp_1$ \& $\Ysp_2$
& (1, 0, 0, 0) &5/12& III
\\
\hline
D3-NS5&D3&9& $\Ysp_1$ \& $\Ysp_2$ & (0, 0, 0, 1)& 4/11 & III
\\
%\hline
\cline{2-7}
&NS5&9& $\Ysp_1$ \& $\Ysp_2$ & (0, 0, 0, 1) &6/13&
III
\\
\hline
D4-NS5&D4&9& $\tilde{\Xsp}$ \& $\Ysp_1$ \& $\Ysp_2$ & (0, 0, 0, 1)& 5/12 &  III
\\
%\hline
\cline{2-7}
&NS5&9& $\tilde{\Xsp}$ \& $\Ysp_1$ \& $\Ysp_2$ & (0, 0, 0, 1)&6/13& III
\\
\hline
D5-NS5&D5&9& $\tilde{\Xsp}$ \& $\Ysp_1$ \& $\Ysp_2$ & (0, 0, 0, 1)& 6/13 &  III
\\
%\hline
\cline{2-7}
&NS5&9&$\tilde{\Xsp}$ \& $\Ysp_1$ \& $\Ysp_2$ & (0, 0, 0, 1)&6/13& III
\\
\hline
D6-NS5&D6&9& $\tilde{\Xsp}$ \& $\Ysp_1$ \& $\Ysp_2$ \& Z & (1, 0, 0, 0)
& 6/13 &  III
\\
%\hline
\cline{2-7}
&NS5&9&$\tilde{\Xsp}$ \& $\Ysp_1$ \& $\Ysp_2$ & (0, 0, 0, 1)& 6/13 & III
\\
\hline
D7-NS5&D7&5& $\Ysp_1$ \& $\Ysp_2$ &(4, 0, 0, 1)&
1/7& III
\\
\cline{2-7}
&NS5&9&$\tilde{\Xsp}$ \& $\Ysp_1$  \& $\Ysp_2$ &(0, 0, 0, 1)&
6/13& III
\\
\hline
D8-NS5&D8&4& $\Ysp_2$ &(5, 0, 0, 1)&1/3& III
\\
\cline{3-7}
&NS5&9&$\tilde{\Xsp}$ \& $\Ysp_1$ &(0, 0, 0, 1)&
6/13& III
\\
\hline
\end{tabular}
}
\label{table_25}
\end{center}
\end{table}

%%%%%%%%%%%%%%%%%%%%%%%%%%%%%%%%%%%%%%%%%%%%%%%%%%%%%%%
%\begin{table}[singI]
\begin{table}[p]
\caption{\baselineskip 14pt
Future singularity of brane worlds in case (I)
with $A_0< 0$.}
\begin{center}
{\scriptsize
\begin{tabular}{|c|c|c|%c||c|c||c||c|
}
\hline
Future singularity&
Type&
Intersecting branes
\\
\hline
\hline
$a=0$  &
D&
D1-D3($\Ysp_1$,$\Zsp$),
D7-D5($\Ysp_1$,$\Zsp$),
D2-D4 ($\Ysp_1$,$\Zsp$),
\\&&
D3-D5 ($\Ysp_1$,$\Zsp$),
D5-D7($\Ysp_1$,$\Zsp$),
D7-D5($\tilde \Xsp$,$\Zsp$)
\\
\cline{2-3}
%Big Crunch
&
M&
M2-M5($\Ysp_1$,$\Zsp$)
\\
\cline{2-3}
%Big Crunch
&
F1&
F1-D3($\Ysp_1$,$\Zsp$)
\\
\cline{2-3}
%Big Crunch
&
NS5&
D3-NS5($\Ysp_1$,$\Zsp$)
\\
%%%%%%%%%%%
\cline{1-3}
$a=\infty$&
D&
D3-D1($\Ysp_2$,$\Zsp$),
D4-D2($\Ysp_2$,$\Zsp$),
D5-D3($\Ysp_2$,$\Zsp$),
D5-D7 ($\tilde \Xsp$,$\Zsp$)
\\
\cline{2-3}
%Big Rip
&
M&
M5-M2($\Ysp_2$,$\Zsp$),
M5-M5($\tilde \Xsp$,$\Zsp$)
\\
%Big Rip
\cline{2-3}
&
F1&
D3-F1($\Ysp_2$,$\Zsp$)
\\
\cline{2-3}
%Big Rip
&
NS5&
NS5-D3($\Ysp_2$,$\Zsp$),
D4-NS5($\tilde \Xsp$,$\Zsp$),
NS5-D4 ($\tilde \Xsp$,$\Zsp$)
\\
\hline
\end{tabular}
}
\label{table_26}
\end{center}
\end{table}
%%%%%%%%%%%%%%%%%%%%%%%%%%%%%%%%%%%%%%%%%%%%%%%%%%%%%%%%%
%\begin{table}[singII]
\begin{table}[p]
\caption{\baselineskip 14pt
Future singularity of brane worlds in case (II)
with $A_0< 0$.
}
%\begin{center}
\hspace{-10mm}
{\scriptsize
\begin{tabular}{|c|c|c|%c||c|c||c||c|
}
\hline
Future singularity&
Type &
Intersecting branes
\\
\hline
\hline
$a=0$  &
 D   &
D3-D1($\Zsp$,$\Ysp_1$),
D3-D1($\Ysp_2$,$\Ysp_1$),
D2-D4($\Ysp_1$,$\Zsp$),
\\
&&
D3-D5($\Ysp_1$,$\Zsp$),
D5-D7 ($\Ysp_1$,$\Zsp$),
D7-D5($\tilde \Xsp$,$\Zsp$)
\\
\cline{2-3}
%Big Crunch
&
M&
M2-M5($\Ysp_1$,$\Zsp$),
M2-M5 ($\Zsp$, $\Ysp_1$),
M5-M2 ($\Ysp_2$,$\Zsp$)
\\
\cline{2-3}
%Big Crunch
&
F1 &
F1-D3($\Ysp_1$,$\Zsp$),
F1-D3 ($\Zsp$,$\Ysp_1$),
D3-F1 ($\Ysp_2$,$\Ysp_1$),
D3-F1 ($\Zsp$,$\Ysp_1$)
\\
\cline{2-3}
%Big Crunch
&
NS5&
D3-NS5($\Ysp_1$,$\Zsp$),
D3-NS5($\Zsp$,$\Ysp_1$),
NS5-D3($\Ysp_1$,$\Zsp$),
NS5-D3($\Ysp_2$,$\Ysp_1$),
\\
&
&
D4-NS5($\Zsp$,$\Ysp_1$),
NS5-D4($\tilde \Xsp$,$\Ysp_1$),
NS5-D4 ($\Zsp$,$\Ysp_1$)
\\
%%%%%%%%%%%
\cline{1-3}
$a=\infty$ &
D &
D5-D7($\tilde \Xsp$,$\Zsp$)
\\
\cline{2-3}
%Big Rip
&
M&
M5-M5($\tilde \Xsp$,$\Zsp$)
\\
\cline{2-3}
%Big Rip
&
F1&
D3-F1($\Ysp_2$,$\Zsp$)
\\
\cline{2-3}
%Big Rip
&
NS5&
NS5-D4 ($\tilde \Xsp$,$\Zsp$)
\\
\hline
\end{tabular}
}
\label{table_27}
%\end{center}
\end{table}
%%%%%%%%%%%%%%%%%%%%%%%%%%%%%%%%%%%%%%%%%%%%%%%%%%%%%%%%%
%\begin{table}[singIII]
\begin{table}[p]
\caption{\baselineskip 14pt
Future singularity of brane worlds in case (III)
with $A_0< 0$.
}
\begin{center}
{\scriptsize
\begin{tabular}{|c|c|c|%c||c|c||c||c|
}
\hline
Future singularity&
Type&
Intersecting branes
\\
\hline
\hline
$a=0$  &
D&
D5-D3($\Ysp_1$,$\Ysp_2$),
D6-D4($\Ysp_1$,$\Ysp_2$)
\\
\cline{2-3}
%Big Crunch
&
NS5&
D6-NS5($\tilde \Xsp$,$\Zsp$),
D5-NS5($\Ysp_1$,$\Ysp_2$),
D7-NS5($\Ysp_2$,$\Ysp_1$),
D7-NS5($\tilde \Xsp$,$\Ysp_1$)
\\
%Big Crunch
&
&
NS5-D3($\Ysp_1$,$\Ysp_2$),
NS5-D4($\Ysp_1$,$\Ysp_2$),
NS5-D5($\Ysp_1$,$\Ysp_2$),
NS5-D7($\Ysp_1$,$\Ysp_2$),
NS5-D7($\tilde \Xsp$,$\Ysp_2$)
\\
%%%%%%%%%%%
\cline{1-3}
$a=\infty$
&
D&
D3-D5($\Ysp_2$,$\Ysp_1$),
D4-D6($\Ysp_2$,$\Ysp_1$),
D3-D7(($\tilde \Xsp$,$\Ysp_2$),$\Ysp_1$)
\\
\cline{2-3}
%Big Rip
&
NS5&
D3-NS5($\Ysp_2$,$\Ysp_1$),
D4-NS5($\Ysp_2$,$\Ysp_1$),
D5-NS5($\Ysp_2$,$\Ysp_1$),
\\
%Big Rip
&
&
NS5-D3($\Ysp_2$,$\Ysp_1$),
NS5-D4($\Ysp_2$,$\Ysp_1$),
NS5-D5($\Ysp_2$,$\Ysp_1$),
NS5-D7($\Ysp_2$,$\Ysp_1$)
\\
\hline
\end{tabular}
}
\label{table_28}
\end{center}
\end{table}
%%%%%%%%%%%%%%%%%%%%%%%%%%%%%%%%

%\begin{table}[pospow]
\begin{table}[p]
\caption{\baselineskip 14pt
Solutions with $d^2\propto h_r^{q-1}$
with $q>1$.
}
\begin{center}
{\scriptsize
\begin{tabular}{|c|c|%c||c|c||c||c|
}
\hline
Case &
Intersecting branes
\\
\hline
\hline
 I   &
D7-D5($\Ysp_2$,$\Zsp$),
D7-D5($\tilde \Xsp$,$\Zsp$)
\\
\cline{1-2}
II&
D3-D1($\Ysp_2$,$\Ysp_1$),
D7-D5($\tilde \Xsp$,$\Zsp$),
D3-F1($\Ysp_2$,$\Ysp_1$),
D3-F1($\Zsp$,$\Ysp_1$),
NS5-D3($\Ysp_2$,$\Ysp_1$),
NS5-D4($\tilde \Xsp$,$\Ysp_1$)
\\
\cline{1-2}
III&
D6-NS5($\tilde \Xsp$,$\Ysp_1$),
D7-NS5($\Ysp_2$,$\Ysp_1$),
D7-NS5($\tilde \Xsp$,$\Ysp_1$),
NS5-D7($\tilde \Xsp$,$\Ysp_2$)
\\
\hline
\end{tabular}
}
\label{table_29}
\end{center}
\end{table}

%%%%%%%%%%%%%%%%%%%%%%%%%%%%%
%\begin{table}[codim1]
\begin{table}[p]
\caption{\baselineskip 14pt
Solutions which give codimension-one brane world models.}
\begin{center}
{\scriptsize
\begin{tabular}{|c||c|c|%c||c|c||c||c|
}
\hline
Case&
Type &
Intersecting branes
\\
\hline
\hline
I  &
(a) &
D7-D5($\Ysp_2$,$\Zsp$),
D5-D7 ($\Ysp_1$,$\Zsp$),
D7-D5($\tilde \Xsp$,$\Zsp$),
D5-D7($\tilde \Xsp$,$\Zsp$)
\\
\cline{2-3} &
(b)&
 no solution
\\
%%%%%%%%%%%
\hline
II &
(a)&
D3-D1 ($\Zsp$,$\Ysp_1$),
D3-D1($\Ysp_2$,$\Ysp_1$),
D3-F1 ($\Ysp_2$,$\Ysp_1$),
D3-F1($\Zsp$,$\Ysp_1$),
\\
&
&
NS5-D3 ($\Ysp_2$,$\Ysp_1$),
NS5-D4($\tilde \Xsp$,$\Ysp_1$),
NS5-D4($\Zsp$,$\Ysp_1$)
\\
\cline{2-3}&
(b)&
D5-D7($\Ysp_1$,$\Zsp$),
D7-D5($\tilde \Xsp$,$\Zsp$),
D5-D7($\tilde \Xsp$,$\Zsp$)
%%%%%%%%%%%%%%%%%%%%%%%%%%%
\\
\hline
III &
(a)&
D7-NS5($\Ysp_2$,$\Ysp_1$),
D7-NS5($\tilde \Xsp$,$\Ysp_1$)
\\
\cline{2-3}&
(b)&
NS5-D7($\Ysp_1$,$\Ysp_2$),
NS5-D7($\tilde \Xsp$,$\Ysp_2$)
\\
\hline
\end{tabular}
}
\label{table_30}
\end{center}
\end{table}
%%%%%%%%%%%%%%%%%%%%%%%%%%%%%%%%%

%%%%%%%%%%%%%%%%%%%%%%%%%%%%%
%\begin{table}[codim1_2]
\begin{table}[p]
\caption{\baselineskip 14pt
The possible future singularities for the codimension-one brane world
for $A_0<0$.
``(In)finite'' means that the brane world reaches the singularity
within a (in)finite proper time.
}
\begin{center}
{\scriptsize
\begin{tabular}{|c||c|c|c|%c||c|c||c||c|
}
\hline
&
Case&
$a=0$ &
$a=\infty$
\\
\hline
\hline
&
I&
D7-D5($\Ysp_2$,$\Zsp$),
D7-D5($\tilde \Xsp$,$\Zsp$)
&
\\
\cline{2-3}
Finite
&
II
&
D3-D1($\Ysp_2$,$\Ysp_1$),
D7-D5($\tilde \Xsp$,$\Zsp$),
D3-F1($\Ysp_2$,$\Ysp_1$),
D3-F1($\Zsp$,$\Ysp_1$)
 &
No solution
\\
&
&
NS5-D3($\Ysp_2$,$\Ysp_1$),
NS5-D4($\tilde \Xsp$,$\Ysp_1$)
&
\\
\cline{2-3}
&
III &
NS5-D7($\tilde \Xsp$,$\Ysp_2$),
D7-NS5($\Ysp_2$,$\Ysp_1$),
D7-NS5($\tilde \Xsp$,$\Ysp_1$)
&
\\
%%%%%%%%%%%
\hline
&
I&
D5-D7($\Ysp_1$,$\Zsp$)
&
D5-D7($\Ysp_2$,$\Zsp$)
%%%%%%%%%%%%%%%%%%%%%%%%%%%
\\
\cline{2-4}
Infinite
&
II &
D5-D7($\Ysp_1$,$\Zsp$),
D3-D1($\Zsp$,$\Ysp_1$),
NS5-D4($\Zsp$,$\Ysp_1$)
&
D5-D7($\tilde \Xsp$,$\Zsp$)
\\
\cline{2-4}
&
III&
NS5-D7($\Ysp_1$,$\Ysp_2$)&
\\
\hline
\end{tabular}
}
\label{table_30_2}
\end{center}
\end{table}

%\vspace{0.1cm}
%%%%%%%%%%%%%%%%%%%%%%%%%%%%%%%%%%%%%%%%%%%%%%%%%%%%%%%%%%%%%%%%%

%%%%%%%%%%%%%%%%%%%%%%%%%%
%\begin{table}[codimn]
\begin{table}[p]
\caption{\baselineskip 14pt
Static solutions which give a
constant effective gravitational coupling
in the limit of $\xi\to \infty$.
}
\begin{center}
{\scriptsize
\begin{tabular}{|c||c|c|c|%c||c|c||c||c|
}
\hline
Case&
$n$ &
Intersecting branes&
$\frac{8\pi G_{\rm eff}}{3}$
\\
\hline
\hline
I  &
3 &
D5-D3($\Ysp_2$,$\Zsp$),
NS5-D3($\Ysp_2$,$\Zsp$)
&
$\frac{\kappa^4\sigma(M_r C-M_s B)^2}
     {3200 \pi^2 M_r^{18/5}M_s^{12/5}}$
\\
\hline
I  &
3 &
D3-D5($\Ysp_1$,$\Zsp$),
D3-NS5($\Ysp_1$,$\Zsp$)&
$\frac{\kappa^4\sigma(M_r C-M_s B)^2}
     {3200 \pi^2 M_r^{12/5}M_s^{18/5}}$
\\
\hline
III&
6&
D3-D7((${\tilde \Xsp}$,$\Ysp_2$),$\Ysp_1$)&
$\frac{\kappa^4\sigma}{18\pi^6M_r^{5/2}}
$
\\
\hline
\end{tabular}
}
\label{table_31}
\end{center}
\end{table}

%T1>Acknowledgments
\section*{Acknowledgments}
K.U. would like to thank H. Kodama, M. Sasaki, and T. Okamura
for continuing encouragement.
The work of N.O. was supported in part by the Grant-in-Aid for
Scientific Research Fund of the JSPS (C) No. 20540283, No. 21$\cdot$09225,
 and (A) No. 22244030.
K.U. is supported by Grant-in-Aid for Young Scientists (B) of JSPS Research,
under Contract No. 20740147.
The authors also acknowledge the long-term workshop
``Gravity and Cosmology 2010'' (YITP-T-10-01)
and the Yukawa International Seminar 2010
``Cosmology --The Next Generation--''(YKIS2010),
held at the Yukawa Institute for Theoretical Physics,
during which a part of this work was performed.

%======================================%
%<<<<<<<<<<<<< REFERENCE >>>>>>>>>>>>>>%
%======================================%

%T1>References

\end{document}